\documentclass[dvipsnames]{aa}

\usepackage{hyperref}
\hypersetup{colorlinks=true, citecolor=PineGreen}
\usepackage[capitalize]{cleveref}
\usepackage[dvipsnames]{xcolor}
\usepackage{xspace}
\usepackage[detect-weight=true, detect-family=true]{siunitx}
\usepackage{booktabs}
\usepackage{multicol}
\usepackage{ulem}
\newcommand{\tconv}{\ensuremath{\tau_\text{conv}}\xspace}
\newcommand{\Ntconv}{\ensuremath{N_{\tconv}}\xspace}
\DeclareSIUnit\coreh{core\textrm{-}h}
\DeclareSIUnit\msol{\textrm{M}_{\odot}}
\DeclareSIUnit\ma{\ma}
\DeclareSIUnit\tconv{\tau_\text{conv}}
\DeclareSIUnit\erg{erg}
\DeclareSIUnit\year{yr}
\newcommand{\figr}[1]{figure~#1}
\newcommand{\figrs}[1]{figures~#1}

\newcommand{\wbing}{well-balancing\xspace}
\newcommand{\leroux}{Cargo--LeRoux\xspace}

\newcommand{\bvf}{BVF\@\xspace}

\newcommand{\dd}{\text{d}}

\newcommand{\ma}{\mathrm{Ma}}
\newcommand{\csound}{\ensuremath{c_\text{sound}}\xspace}

\newcommand{\ausmpup}{\ensuremath{\text{AUSM}^{+}\text{-up}}\xspace}
\newcommand{\ausmpups}{{\tiny \ausmpup}}
\newcommand{\apup}{\ensuremath{\text{A}^{+}\text{-up}}}
\newcommand{\ausmp}{\ensuremath{\text{AUSM}^{+}_\text{B}\text{-up}}\xspace}
\newcommand{\ausmps}{{\tiny \ausmp}}
\newcommand{\aup}{\ensuremath{\text{A}^{+}_\text{B}\text{-up}}}
\newcommand{\roe}{Roe\xspace}

\newcommand{\slh}{SLH\@\xspace}
\newcommand{\mesa}{MESA\@\xspace}
\newcommand{\fcbm}{\ensuremath{f_{\text{\cbm}}}}

\newcommand{\sht}{\textit{shtools}\xspace}
\newcommand{\supad}{\ensuremath{\Delta\nabla}\xspace}
\newcommand{\specie}[2]{\ensuremath{^{#1}}{#2}\xspace}
\newcommand{\rans}{RA-ILES\xspace}
\newcommand{\iles}{ILES\xspace}
\newcommand{\nekin}{\ensuremath{\mathcal{N}_{\epsilon_k}}\xspace}
\newcommand{\rib}{\ensuremath{\text{Ri}_{\text{B}}}\xspace}
\newcommand{\vsub}[1]{\ensuremath{v_{\text{#1}}}\xspace}
\newcommand{\ve}{\vsub{e}}
\newcommand{\rms}{rms\xspace}
\newcommand{\vrms}{\vsub{rms}}
\newcommand{\vds}{\ensuremath{v_{\Delta s}}\xspace}
\newcommand{\mrms}{\ensuremath{\ma_\text{rms}}\xspace}
\newcommand{\rczb}{\ensuremath{r_\text{CZ,0}}\xspace}
\newcommand{\rczt}{\ensuremath{r_\text{CZ,1}}\xspace}
\newcommand{\fapdiff}{{\ensuremath{f_a^\text{p}}}\xspace}
\newcommand{\fa}{{\ensuremath{f_a}}\xspace}
\newcommand{\res}[2]{\ensuremath{\num{#1}\times\num{#2}^2}}
\newcommand{\lres}{\res{180}{90}\xspace}
\newcommand{\mres}{\res{360}{180}\xspace}
\newcommand{\mrest}{\res{360}{240}\xspace}
\newcommand{\hres}{\res{540}{360}\xspace}
\newcommand{\vhres}{\res{810}{540}\xspace}
\newcommand{\eos}{EoS\xspace}
\newcommand{\ci}{C+17\xspace}
\newcommand{\cii}{C+19\xspace}
\newcommand{\mlt}{\text{MLT}\xspace}
\newcommand{\cbm}{\text{CBM}\xspace}
\newcommand{\tdiff}{\ensuremath{\tau_\text{diff}}\xspace}
\newcommand{\tofn}[1]{\ensuremath{t\left(\Ntconv=#1\right)}\xspace}
\newcommand{\ncz}{\ensuremath{\#_{\text{CZ}}}\xspace}
\newcommand{\vpc}{\ensuremath{u_{\text{cell}}}}
\newcommand{\pec}{P\'eclet\xspace}

\newcommand\tspace{\rule{0pt}{2.6ex}}
\newcommand\bspace{\rule[-1.2ex]{0pt}{0pt}}

\begin{document}

\title{Multidimensional low-Mach number time-implicit hydrodynamic simulations of convective helium shell burning in a massive star}
\author{L. Horst\inst{1}\and
        R. Hirschi\inst{2,3}\and
        P. V. F. Edelmann\inst{4}\and
        R. Andr\'assy\inst{1}\and
        F. K. R\"opke\inst{1,5}}

\titlerunning{Helium shell burning}
\authorrunning{Horst et al.}

\institute{%
  Heidelberger Institut f\"{u}r Theoretische Studien,
  Schloss-Wolfsbrunnenweg 35, 69118 Heidelberg, Germany\\
  \email{leonhard.horst@posteo.de}
  \and Astrophysics Group, Keele University, Keele, Staffordshire ST5 5BG, UK
  \and Kavli Institute for the Physics and Mathematics of the Universe (WPI), University of Tokyo, 5-1-5 Kashiwanoha, Kashiwa 277-8583, Japan
  \and
  X Computational Physics (XCP) Division and Center for Theoretical Astrophysics (CTA),
  Los Alamos National Laboratory,
  Los Alamos,
  NM 87545,
  USA
  \and
  Zentrum f\"ur Astronomie der Universit\"at Heidelberg,
  Institut f\"ur Theoretische Astrophysik,
  Philosophenweg 12,
  69120 Heidelberg,
  Germany
}

\date{Received 17 March 2021; accepted 30 June 2021}
\abstract{%
  \textit{Context.} A realistic parametrization of convection and convective
  boundary mixing in conventional stellar evolution codes is still the subject
  of ongoing research. To improve the current situation, multidimensional
  hydrodynamic simulations are used to study convection in stellar interiors.
  Such simulations are numerically challenging, especially for flows at low
  Mach numbers which are typical for convection during early evolutionary
  stages.
  \\
  \textit{Aims.} We explore the benefits of using a low-Mach hydrodynamic flux
  solver and demonstrate its usability for simulations in the astrophysical
  context. Simulations of convection for a realistic stellar profile are
  analyzed regarding the properties of convective boundary mixing.
  \\
  \textit{Methods.} The time-implicit Seven-League Hydro (\slh) code was used to
  perform multidimensional simulations of convective helium shell burning based
  on a \SI{25}{\msol} star model. The results obtained with the low-Mach
  \ausmpup solver were compared to results when using its non low-Mach variant
  \ausmp. We applied \wbing of the gravitational source term to maintain the
  initial hydrostatic background stratification. The computational grids have
  resolutions ranging from \lres to \vhres cells and the nuclear energy release
  was boosted by factors of \numlist{3e3;1e4;3e4} to study the dependence of the
  results on these parameters.
  \\
  \textit{Results.} The boosted energy input results in convection at Mach
  numbers in the range of \num{e-3} to \num{e-2}. Standard mixing-length theory
  (\mlt) predicts convective velocities of about $\num{1.6e-4}$ if no boosting
  is applied. The simulations with \ausmpup show a Kolmogorov-like inertial
  range in the kinetic energy spectrum that extends further toward smaller
  scales compared with its non low-Mach variant. The kinetic energy dissipation
  of the \ausmpup solver already converges at a lower resolution compared to
  \ausmp. The extracted entrainment rates at the boundaries of the convection
  zone are well represented by the bulk Richardson entrainment law and the
  corresponding fitting parameters are in agreement with published results for
  carbon shell burning. However, our study needs to be validated by simulations
  at higher resolution. Further, we find that a general increase in the entropy
  in the convection zone may significantly contribute to the measured
  entrainment of the top boundary.
  \\
  \textit{Conclusion.} This study demonstrates the successful application of
  the \ausmpup solver to a realistic astrophysical setup. Compressible
  simulations of convection in early phases at nominal stellar luminosity will
  benefit from its low-Mach capabilities. Similar to other studies, our
  extrapolated entrainment rate for the helium-burning shell would lead to an
  unrealistic growth of the convection zone if it is applied over the lifetime
  of the zone.  Studies at nominal stellar luminosities and different phases of
  the same convection zone are needed to detect a possible evolution of the
  entrainment rate and the impact of radiation on convective boundary mixing.
  }
\maketitle

\section{Introduction}
\label{sec:intro}

\begin{figure}
  \centering
  \includegraphics[width=\columnwidth]{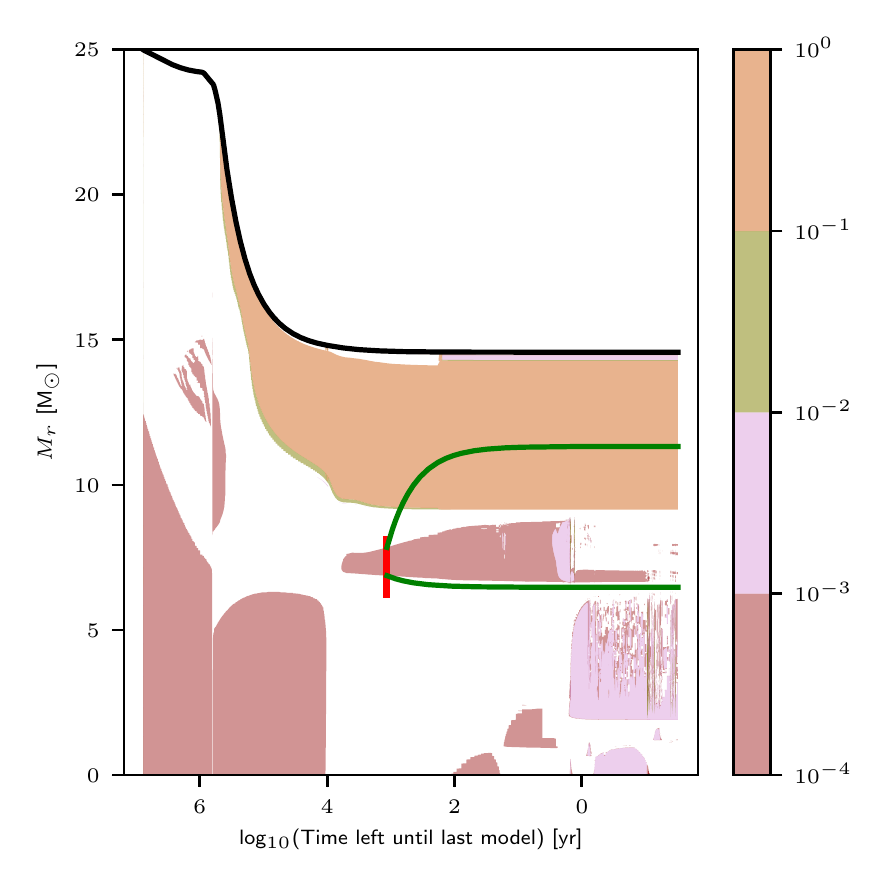}
  \caption{Convective regions during the evolution of an 1D \SI{25}{\msol} star
  model simulated with the \mesa code.  Shaded regions correspond to convection
  zones. The color-shading represents the \mlt-predicted Mach number. The black
  solid line denotes the total mass of the model. The red vertical line
  indicates the point in the evolution at which the \slh simulations start and
  the mass extent of the initial model. The green lines indicate the
  mass entrainment at the upper and lower boundaries as extracted from the
  3D hydrodynamic simulations. See discussion in \cref{sec:cbm}.}
  \label{fig:kipp_initial}
\end{figure}

Mixing induced by convection in the stellar interior plays an essential role in
the evolution of stars. Parametrizing its complex multidimensional nature in
one-dimensional (1D) stellar evolution codes is, however, particularly
difficult. A reliable prescription of convective effects in 1D
codes is still lacking today and the resulting stellar evolution models depend on the
specific choice of the employed paramterization and the particular parameter
values. This is, for example, demonstrated in recent studies by
\citet{kaiser2020a} and \citet{davis2019a} on uncertainties in core properties
and nucleosynthesis for massive stars.

With asteroseismic data of observed stars, it is possible to determine
properties of the stellar interiors (see \citealp{aerts2021a} for a detailed
review). They can be utilized to narrow down the range of possible parameters.
It is now possible to determine that some convective boundary mixing models
provide a better fit to certain asteroseismic observations than others
\citep[e.g.,][]{viani2020a,angelou2020a,michielsen2019a,pedersen2018a,pedersen2021a},
but probing the small-scale physics of the mixing is beyond the reach of even
state-of-the-art asteroseismology.

A complimentary approach to improve the current situation is to perform
multidimensional simulations by numerically solving the equations of fluid
dynamics for realistic stellar models. In such simulations, convection develops
self-consistently and their detailed analysis provides insights into the
fundamental processes at play. This way, currently used parametrizations of
convection in 1D codes can be constrained or discarded and new prescriptions
may be developed.

The complex problem of convection and associated mixing of material across the
interfaces into stable zones in the stellar context is subject of active, ongoing
research.  Numerical simulations become particularly challenging when the flow of
interest is slow compared to the speed of sound, that is for small values of
the Mach number
\begin{equation}
  \ma = \frac{v}{\csound},
  \label{eq:machnumber}
\end{equation}
where $v$ is the flow velocity and \csound is the local speed of sound. One
challenge is the restricted step size of conventional explicit time stepping
schemes which must be smaller than the sound crossing time of a single grid
cell for numerical stability. Thus, at low Mach numbers, an excessively large
number of time steps is needed to evolve the slow flow and explicit schemes
become inefficient.  Additionally, artifacts of the numerical discretization
must be kept at a very low level because inaccuracies can quickly lead to
spurious velocities at the same order as the flow of interest. Hence,
appropriate numerical techniques must be chosen carefully.

For massive stars, low-Mach number flows typically arise in convection during
the early phases of stellar evolution, see for example the evolution of the
\SI{25}{\msol} star depicted in \cref{fig:kipp_initial}.  Inaccuracies in the
1D prescription of convection in these phases propagate to all subsequent
evolutionary phases and also enter predictions for the final stages of stars
and observables. We therefore believe that successful simulations of these
challenging settings are crucial to further improve the agreement between
stellar modeling and observations.

One approach to meet the challenges of low-Mach flows is to modify the
underlying hydrodynamic equations. This is, for example, done in the MAESTRO
code \citep{almgren2007a,nonaka2010a,fan2019a}, where the Euler equations are
modified to exclude the physics of sound waves and to ensure the correct
scaling of leading-order terms in the low-Mach limit. This permits larger time
steps and increases the efficiency for slow flows. An example for low-Mach
simulations with the MAESTRO code are the three-dimensional (3D) simulations of
core hydrogen burning by \citet{gilet2013a}. Another approach is to perform
implicit time stepping while solving the unmodified Euler equations, including
sound waves.  The time step size is then only restricted by the desired
accuracy at which the flow is to be followed. This is for example employed by
the MUSIC code in combination with a staggered spatial grid. Benchmark tests
have shown that the code is able to evolve flows at Mach numbers down to
$\ma\approx\num{e-6}$ \citep{viallet2016a} and that a hydrostatic atmosphere
remains stable \citep{goffrey2017a}.

The Seven-League (\slh) hydro code, which is used for the simulation presented
here, is designed to tackle the numerical difficulties of low-Mach flows. It
uses implicit time stepping and solves the fully compressible Euler equations.
Furthermore, it applies special numerical flux functions with enhanced low-Mach
capabilities in combination with \wbing techniques to improve the
representation of slow flows. This way, the \slh code is able to capture flows
at low and moderate Mach numbers on the same grid.

The work we present in this paper aims at contributing to the recent effort to
improve the understanding of the complex behavior of convection by means of
hydrodynamic simulations. We demonstrate the benefits from using the
low-Mach-number flux \ausmpup even at moderate Mach numbers. For this, 3D
simulations of convective He-shell burning in a \SI{25}{\msol} star are
presented and analyzed regarding general properties of the turbulent
convection. In addition, we complement recent efforts to characterize
convective boundary mixing by means of multidimensional simulations
\citep[e.g.,][]{meakin2007b,woodward2015a,
jones2017a,cristini2017a,cristini2019a,pratt2017a,pratt2020a,higl2021a}.

The paper is structured as follows: In \cref{sec:slh} we briefly describe the
basic properties of the \slh code. In \cref{sec:setup} the initial conditions
for the  simulations are presented along with a detailed description of mapping
the 1D model to the \slh grid and the applied energy boosting. In the 1D and
two-dimensional (2D) test simulations presented in \cref{sec:prelimtests} we
assess the hydrostatic stability of the initial profile using the \leroux\wbing
method and determine the required amount of artificial energy boosting. The
corresponding 3D simulations are analyzed in \cref{sec:3Dresults} regarding
properties of the turbulent convective flow and boundary mixing.
\cref{sec:conclusion} summarizes the results.

\section{The Seven-League Hydro (\slh) code}
\label{sec:slh}

The hydrodynamic simulations presented in this paper are performed with the
\slh code \citep{miczek2013a,edelmann2014a}. It solves the fully compressible
Euler equations in a finite volume approach. The underlying equations are
formulated in general, curvilinear coordinates and mapped onto a logically
rectangular computational grid, following the method
of \citet{kifonidis2012a}.  This allows one to construct almost arbitrary grid
geometries that can be adapted to the physical setup that is investigated
\citep{miczek2013a}. The Helmholtz equation of state (\eos) \citep{timmes2000a}
is implemented and accounts for radiation pressure and degeneracy effects. The
hydrodynamic equations are coupled to a nuclear reaction network
\citep{edelmann2014a}.

The \slh code is designed to simulate hydrodynamic phenomena in the context of
stellar astrophysics for flows at low and intermediate Mach numbers.  The
following discussion briefly summarizes how the numerical challenges,
especially for low-Mach flows, are approached in \slh. For a more in-depth
description of the applied methods we refer the reader to
\citet{edelmann2021a}, \citet{edelmann2014a}, and \citet{miczek2013a}.

\subsection{Flux Solver}

\citet{miczek2015a} and \citet{barsukow2017a} demonstrated that low-Mach flows
require special numerical flux functions because common schemes, as for example
the popular Roe solver \citep{roe1981a}, suffer from excessive numerical
dissipation. A variety of flux functions with improved low-Mach capabilities
can be found in the literature. One promising method that seems to be
applicable to problems in stellar astrophysics is the \ausmpup scheme
\citep{liou2006a} which is implemented into \slh with a slight modification. As
described in \citet{edelmann2021a}, the \slh implementation uses two
independent parameters to control the velocity diffusion (\fa) and pressure
diffusion (\fapdiff), respectively. The original \ausmpup scheme only uses a
single parameter. It has been demonstrated by \citet{horst2020a} that compared
to the classical \roe scheme the \ausmpup solver significantly improves the
accuracy at which internal gravity waves can be followed for group velocities
at low Mach numbers. For all simulations with the \ausmpup solver presented in
this paper, we set the parameters to the values $\fapdiff=\num{0.1}$
and $\fa=\num{e-10}$, which has proven to yield robust results in previous test
simulations.

In \cref{sec:3Dresults} we compare simulations with the \ausmpup solver with
its basic variant \ausmp in order to demonstrate the improved results when
using \ausmpup. The \ausmp scheme is a subclass of the \ausmpup scheme and is
obtained by disabling the scaling of the incorporated velocity and pressure
diffusion with Mach number. This scaling ensures the correct behavior of
leading terms of the pressure field in the limit of $\ma\to0$
(\citealp[see][Sec.~3.2]{liou2006a} for details). Hence, \ausmp does not have
enhanced low-Mach capabilities. In \slh, the \ausmp solver option is obtained
by setting $\fapdiff=\fa=1$.

\subsection{Well-balancing}

Maintaining hydrostatic equilibrium is not trivial in finite volume codes
because commonly gravity is discretized differently than the conserved
variables and enters the equations in an operator-split approach. Hence, even
if the initial data on the computational grid is formally in perfect
hydrostatic equilibrium, a residual source term in the momentum and energy
parts of the Euler-equations will remain \citep[see,
e.g.,][]{kappeli2016a,popov2019a,berberich2021a,edelmann2021a}. For the \slh
code, \citet{edelmann2021a} demonstrate that proper \wbing techniques allow to
simulate convection at $\ma\sim\num{e-4}$. However, this requires methods that
have become available only after the simulations of helium shell burning were
carried out. In the simulations presented here, we use the multidimensional
extension \citep{edelmann2021a} of the 1D \leroux\wbing scheme
\citep{cargo1994a}. \citet{edelmann2021a} show that it is not possible to
perform simulations at Mach numbers considerably smaller than
$\ma\sim\num{e-3}$. At Mach numbers below this threshold the flow is
deteriorated by discretization errors. Thus, for our study, the energy
generation from helium burning that drives the convection has to be boosted by
three orders of magnitude to increase the convective velocities, see
\cref{sec:boosting} and \cref{sec:2Dtesting}. Still, \leroux\wbing is crucial
to maintain the background stratification, as demonstrated in
\cref{sec:testCL}.

\subsection{Time stepping}

To circumvent the small time step sizes of explicit time marching schemes, the
\slh code applies implicit time stepping. Here, the time step size is not
restricted by numerical stability requirements but only by the desired accuracy
at which the flow is to be followed. At low Mach numbers, the large time steps
and hence smaller number of total steps overcompensates the higher
computational costs of a single step compared with explicit schemes. For the
simulations presented in this paper the ESDIRK23 scheme \citep{hosea1996a} is
used, which is second order accurate in time. The resulting system of
nonlinear equations is solved with the Newton-Raphson method.

For all simulations presented here, linear reconstruction is used.  Slope
limiter are usually required to diminish oscillations at steep gradients.
However, the partially discontinuous spatial derivatives of common limiters
deteriorate the convergence rate of the Newton-Raphson method. Further tests
are needed to explore their possible applications in implicit \slh simulations.

\section{Model setup}
\label{sec:setup}

\subsection{Construction of the initial model}

The initial conditions for the hydrodynamic \slh simulation are based on an 1D
model obtained with the open-source stellar evolution code \mesa
\citep{paxton2011a, paxton2013a, paxton2015a, paxton2019a},

The model corresponds to a \SI{25}{\msol} star at solar metallicity ($Z=0.014$)
evolved until the exhaustion of core oxygen burning.  The model develops a
convective helium burning shell (at log$_{10}$(time left)$ \lesssim 4$ in
Fig.\,\ref{fig:kipp_initial}) following core helium burning. The numerical
settings are similar to \citet{kaiser2020a} (see their section 3) and briefly
summarized here. Convective zones are determined using the Schwarzschild
criterion, which neglects chemical gradients. It states that regions are
convective if the superadiabaticity $\supad$ is positive, that is
\begin{align}
  \supad := \nabla - \nabla_\text{ad} > 0,
  \label{eq:schwarz}
\end{align}
where $\nabla_\text{ad}$ denotes the adiabatic temperature gradient while the
actual temperature gradient of the gas is given by $\nabla = \dd \ln T / \dd
\ln P$.

Convection is parametrized using \mlt and a mixing length of $\ell_{\mlt} =
1.6\, H_P$, where $H_P$ denotes the pressure scale height
\begin{align}
  H_P = -\frac{\dd r }{\dd P} P.
  \label{eq:hp}
\end{align}
To model convective boundary mixing (\cbm), the exponentially-decaying
diffusion approach of \citet{freytag1996a} and \citet{herwig1997a} is used. The
corresponding diffusion coefficient is \citep{herwig1997a}:
\begin{equation}
  D_{\text{\cbm}} = D_0(f_0) \, \exp\left( \frac{-2 \left[r-r_0(f_0)\right]}{\fcbm \, H^{\rm{CB}}_{P}} \right),
  \label{eq:expD}
\end{equation}
where the free parameter $f_\text{\cbm}$ determines the extent of the \cbm in
terms of the pressure scale height at the boundary of the convection zone
$H^{\rm{CB}}_{\rm{P}}$. For the top boundary of convective regions, $D_0(f_0)$
is the \mlt diffusion coefficient evaluated at $r_0(f_0) = r_{\text{CB}} -
(f_0\,H_p)$ where $r_\text{CB}$ is the radius of the boundary as given by
\cref{eq:schwarz}. The free parameter $f_0$ ensures that the diffusion
coefficient is calculated inside the convection zone to avoid the sharp drop in
$D$ in the immediate vicinity of the boundary. The diffusion coefficient
$D_{\text{\cbm}}$ is applied for radii larger than $r_0(f_0)$ until it drops
below \SI{e2}{\centi\meter\per\second}.
For the bottom boundary of convective regions, the scheme is adapted to apply
\cbm below the convective boundary. The initial 1D model was obtained with the
parameters $\fcbm = 0.022$ and $0.0044$ for the top and bottom boundaries,
respectively.  At both boundaries, $f_0 = 0.025$ was used. We refer to
\citet{kaiser2020a} for a discussion of these parameters and the related
uncertainties.

Our simulations focus on convection in the helium-burning shell. The red
vertical line in \cref{fig:kipp_initial} indicates the evolutionary point at
which the \slh simulations start and the extent in mass coordinates of the
simulation domain. It corresponds to the early phase of helium-shell burning
when the radial extent of the shell is still relatively small. Compared with
later phases, this enables a better resolution at convective boundaries for a
fixed computing budget. Choosing the convective shell also allows us to study
two boundaries rather than only one for convective cores.

Models from stellar evolution codes typically exhibit step-like transitions in
the 1D profiles, for example in the profiles of species abundances or
thermodynamic quantities such as entropy. Even with moderate \cbm
parameters, such as used in the 1D input models, convective boundaries are very
narrow. This is problematic for conventional hydrodynamic simulations because,
if possible at all, a high number of grid cells is necessary to spatially resolve
such transitions. Furthermore, we found in preliminary 2D test simulation that
the steep gradients lead to strong initial flows in the convection zone. This
effect was diminished, yet not fully resolved, for low resolution runs by
applying rather strong smoothing to the initial profiles.

\slh simulations require the initial conditions to accurately fulfill the
equation of hydrostatic equilibrium. This is not guaranteed for the 1D input
profiles after they have been smoothed. Therefore, the equation of hydrostatic
equilibrium needs to be integrated again while prescribing the profile of one
thermodynamic quantity from the 1D \mesa profiles. It is important that in this
process the convection zone, characterized by a negative Brunt-V\"ais\"al\"a
frequency (\bvf), is maintained. For a nonrotating star, the \bvf is given by
\citep[e.g., see][Sect.  6.4.1]{maeder2009a}
\begin{align}
  N^2 = \frac{g\delta}{H_P}\left(\nabla_\text{ad}-\nabla + \frac{\varphi}{\delta}\nabla_\mu \right),
  \label{eq:bvf}
\end{align}
where $g$ is the magnitude of the gravitational acceleration, $\delta =
-(\partial\ln\rho / \partial \ln T)_{P,\mu}$,  $\varphi =
(\partial\ln\rho/\partial\ln\mu)_{P,T}$, and the gradient in mean molecular
weight $\mu$ reads $\nabla_\mu=\dd\ln\mu/\dd\ln P$. These quantities are determined by the
\eos. For the simulations presented here, we follow the approach of
\citet{edelmann2017a} to reproduce a given 1D profile of the superadiabaticity
\supad. This allows one to directly control the extent of the convective region in
the initial condition if the chemical gradient can be neglected in the
convection zone, as is the case for our model. It also ensures that the value
of \supad is reasonably close to zero and does not lead to an initial
convective flow that is mainly driven by an excess in superadiabaticity.
We construct the initial model on a radial grid that is much finer than the
actual computational grid in \slh. For the computational grid, the initial state is
obtained by interpolating from the fine grid to the positions of the respective
cell centers. Because of the fine input grid, interpolation errors are
negligibly small.

Preliminary \slh simulations revealed that setting the superadiabaticity on
the \slh grid to a value of \num{-1.5e-5} whenever $\supad_\text{\mesa} >
\num{-e-3}$ leads to a gentle transition from the initial hydrostatic
stratification to fully developed convection and avoids a large initial peak in
kinetic energy at the onset of convection. The slightly stable stratification
would considerably hinder convection if the nominal luminosity was used.
Because we have to increase the energy input anyway, this is not an issue for
the simulations presented here.

In addition to the convective shell, parts of the radiative zones which lie
above and below the convection zone are included in the computational domain.
Their respective radial extent is chosen to be one half of the extent of the
convection zone itself. This way, the impact of the top and bottom boundary
conditions will be reduced while keeping the computational cost at a moderate
level.

\begin{figure}
  \centering
  \includegraphics[width=\columnwidth]{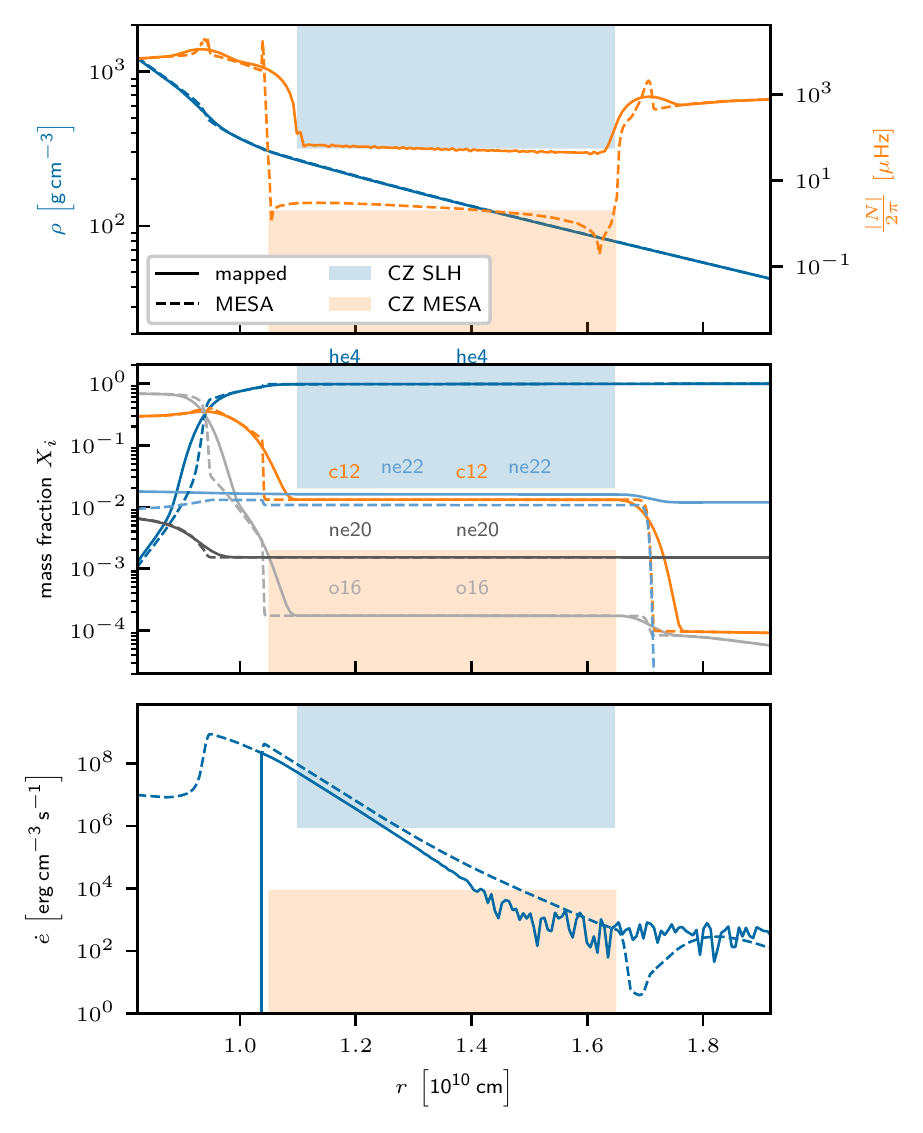}
  \caption{Initial profiles for the underlying 1D \mesa model (dashed) and the
  mapped \slh model (solid lines). The shaded areas mark the convection zone for mapped
  profiles (blue) and \mesa profiles (orange). The oscillatory behavior of the
  energy generation is a numerical artifact at negligible amplitudes.}
  \label{fig:line_initprofs}
\end{figure}
The resulting profiles after smoothing and mapping the region of interest from
the 1D stellar model to the computational grid are shown in
\cref{fig:line_initprofs}. The density closely follows the profiles as given by
the 1D \mesa input model. However, smoothing changes the profile of the \bvf
and alters the size of the convection zone (shaded areas
in \cref{fig:line_initprofs}). Especially the position of the bottom boundary
changes and the convection zone starts at a somewhat larger radius in the
mapped model.

In \cref{sec:cbm} we measure the mass entrainment across the boundaries of the
convective zone. An often employed quantity to characterize the resistance to
such a mixing (also called stiffness) is the bulk Richardson number $\rib$. For
comparability, we follow the notation of \citet{cristini2017a, cristini2019a}
(\ci and \cii hereafter) and write
\begin{align}
  \rib = \frac{\Delta B\, l}{\vrms^2},
  \label{eq:rib}
\end{align}
where \vrms is the \rms velocity of the convection and the integral
length scale $l$ of the convection is set to one half of the pressure scale
height at the boundary. The buoyancy jump $\Delta B$ is given by
\begin{align}
  \Delta B = \int_{r_c-\Delta r}^{r_c+\Delta r} N^2\, \dd r,
  \label{eq:bjump}
\end{align}
where $r_c$ is the radial position of the respective boundary. The integration
width $\Delta r$ is a somewhat arbitrary parameter but should be chosen such
that it includes the full region of the evanescent convective flow at the
boundaries. Following \cii, we set $\Delta r$ to a quarter of the local
pressure scale height. Compared to our measurements of the boundary widths
given in \cref{sec:bwidth} (see \cref{tab:widthboost}), this seems to be an
appropriate value for the top boundary but might overestimate the bottom
boundary.

The definition of \ci for $l$ and $\Delta r$ is not applicable for convection
zones that are thinner than a pressure scale height and some form of
correlation function of the turbulent flow field might be more appropriate.
This, however, is not easily obtained in 1D stellar evolution codes. Generally,
the definition of $l$ and $\Delta r$ is ambiguous in the astrophysical
literature which makes it difficult to directly compare the values of the bulk
Richardson number in simulations carried out by different groups
\citep[see for example][]{meakin2007b, arnett2009a, salaris2017a, cristini2017a,
collins2018a, higl2021a}.

To assess the impact of the applied smoothing on the stiffness, we compare the
numerator  of \cref{eq:rib} for the original 1D \mesa input model and the
mapped \slh model at the respective top and bottom boundary. We find
\begin{align}
  \left.\frac{(\Delta B\, l)_\text{\mesa}}{(\Delta B\, l)_\text{\slh}}\right|_{\text{bot}} \approx 2.9,\quad
  \left.\frac{(\Delta B\, l)_\text{\mesa}}{(\Delta B\, l)_\text{\slh}}\right|_{\text{top}} \approx 1.6,
  \label{eq:1dstiff}
\end{align}
which indicates that the mapping only has a moderate impact on the stiffness of
the boundary.

Due to the computational costs involved, not all nuclear species of the \mesa
nuclear network can be included to the \slh simulation.  Instead, we only
account for \specie{4}{He}, \specie{12}{C}, \specie{16}{O}, \specie{20}{Ne},
and \specie{22}{Ne}. The abundance profile of each species, except for
\specie{22}{Ne}, is taken directly from the \mesa model and smoothed in the
same way as the other input profiles. The abundance of \specie{22}{Ne} follows
from the condition $\sum_i X_i = 1$ in every cell, where $X_i$ is the mass fraction
of species $i$. The resulting profiles are shown in the middle panel of
\cref{fig:line_initprofs}.

Although the smoothing procedure causes the \slh model to slightly deviate from
the original 1D \mesa model, the \mesa model involves uncertainties of its own.
Therefore, we still consider the \slh model to be representative of typical
conditions expected in He-burning shells of massive stars.

\subsection{Energy generation and boosting}
\label{sec:boosting}

The energy release is calculated using the JINA REACLIB reaction files
\citep{cyburt2010a} and displayed in the lowest panel of
\cref{fig:line_initprofs}. From the profile of the energy generation rate as
given directly by the \mesa model (dashed line) it is apparent that the peak of
nuclear burning does not coincide with the convection zone (orange shade) but
instead is located somewhat beneath. This is common for burning shells, which
develop convection above the energy peak where the temperature gradient becomes
steeper than the adiabatic one.

To ensure that convection is driven by the actual energy input and not by
numerical artifacts, the nominal energy input must be boosted. The strength of
the required boosting is determined in \cref{sec:2Dtesting}. We couple the
boosting of the energy generation to the abundance of \specie{4}{He} such that
only regions are boosted where the mass fraction of \specie{4}{He} is higher
than \SI{90}{\percent} of the initial abundance in the convection zone, that is for
abundances higher than \num{0.87}. The energy input is turned off everywhere
else.

\subsection{Thermal diffusion}
\label{sec:thermaldiff}

Thermal radiation is treated in the diffusion limit in \slh. This is justified
by the high optical depth in the interior regions of stars. While the 1D input
profile from the MESA code is in thermal balance, that is the energy flux equals
the integrated energy generation, this is not true anymore within the
convection zone of the \slh simulations with boosted energy generation.
Radiative effects certainly are crucial over the long timescales covered in
simulations of stellar evolution. However, for the much shorter dynamical
timescales we expect the imbalance to be of minor importance: Following the
same arguments as \citet{horst2020a}, we calculate the thermal adjustment
timescale \citep[e.g.,][Sect.  3.2.]{maeder2009a} via
\begin{align}
  \tdiff(\Delta x_{\mathrm{diff}}) \sim \frac{(\Delta x_{\mathrm{diff}})^{2}}{K},\quad
  K = \frac{4\,a\,c_{\mathrm{light}}\,T^3}{3\,\kappa\,\rho^2\,C_{\mathrm{P}}},
  \label{eq:tdiff}
\end{align}
where $\Delta x_{\mathrm{diff}}$ is a typical diffusion length scale, the
radiation constant $a =
\SI{7.57e-15}{\erg\per\cubic\centi\meter\per\kelvin\tothe{4}}$ and $C_\text{P}$
denotes the specific heat of the gas at constant pressure. All other values
have their usual meanings. The opacity $\kappa$ that enters the thermal
diffusivity $K$ is taken from the 1D MESA profile. Because advective and
diffusive processes have a different temporal and spatial scaling, it is not
clear how to scale the opacity with our energy boosting. Therefore, we keep the
opacity at its stellar value in this study.

Assuming as typical length scale the radial grid spacing of the finest
resolution that will be used (\num{810} radial cells, see
\cref{sec:3Dresults}), we find a mean adjustment timescale of
$\overline{\tdiff}(\delta r_{810}) = \SI{5e2}{\hour}$. The timescale is
shortest at the outermost regions where the opacity is the smallest, but is
always larger than \SI{e2}{\hour} (see \cref{fig:line_tdiff}). This is
at the order of our longest runs, which, however, have lower radial
resolution than what is assumed in this estimate. Taking the convection zone as
typical length scale we obtain $\overline{\tdiff}(\delta r_\text{CZ}) \approx
\SI{4e7}{\hour} \approx \SI{4.5e3}{\year}$ which is orders of magnitude longer
than all of our simulations. We therefore conclude that for the particular
simulations presented here the effect of thermal diffusion is negligible and
that the thermal imbalance due to our increased energy input does not impact
the global structure of the star over the duration of our simulations.

\subsection{Discretizing the computational domain}
\label{sec:discret}

To reduce the computational costs, we have to use a spherical-wedge grid
geometry, although this choice eliminates the large-scale flows seen in
comparable $4\pi$ simulations of \citet{woodward2015a}, \citet{jones2017a},
\citet{andrassy2020a}, and \citet{gilet2013a}. The chosen wedge geometry
reduces the computational cost by a factor of \num{32} compared to a full
$4\pi$ simulation at the same vertical and horizontal resolution.

For the $4\pi$ simulations mentioned above, we calculate the aspect ratios (we
adopt here the formulation of \citealp{jones2017a}) as
$(r_\text{top}-r_\text{bot})/r_\text{top}$, where $r_\text{top}$,
$r_\text{bot}$ denote the radial position of the bottom and top boundary of the
convection zone. The aspect ratio for the He-flash simulation of
\citet{woodward2015a} is about \num{0.67}, for the O-shell simulation of
\citet{jones2017a,andrassy2020a} it is about \num{0.5} and for the H-core
burning of \citet{gilet2013a} it is \num{1}. With decreasing aspect ratio, the
maximum possible size of convection cells decreases and so does the impact of
restricting the flow to a spherical wedge. The He-shell simulation presented
here has an aspect ratio of only \num{0.32}. Furthermore, the study by
\citet{gilet2013a} indicates that, while the flow morphology differs distinctly
between their hydrogen core simulations for full $4\pi$ and single octant
domains, basic turbulent properties and mixing rates are in a reasonable
agreement. From this, we expect that the imprint of the restricted geometry on
our results is sufficiently small. However, the influence of the domain size
should be assessed in more detail in future studies.

We set the horizontal extent of the computational domain to be as twice as
large as the vertical extent of the convection zone. This enables the formation
of two large vortices, a typical phenomenon we observe in 2D and 3D
simulations. The corresponding opening angle is about $\SI{45}{\degree}$. The
constant grid spacing giving cell aspect ratios ranging from roughly unity at
the bottom to one half at the top of the domain.

The lowest resolution that will be used in this study has \num{180} vertical
cells and \num{90} horizontal cells. This ensures that the pressure scale
height is resolved by at least \num{25} cells and that the initial transitions
from radiative to convective regions as given by the profile of the \bvf are
resolved by at least \num{20} cells.

Periodic boundary conditions are employed in horizontal direction. In both
radial directions, layers of two cells are added (ghost cells). They are
initialized with the mapped hydrostatic state but are not evolved in time.

\section{1D and 2D test simulations}
\label{sec:prelimtests}

While proper turbulent behavior of convection can only be followed in 3D
simulations, much cheaper 1D and 2D simulations are well suited to test
stability and basic properties of the initial hydrostatic stratification. Such
low-dimensional simulations are utilized in this section to demonstrate that
applying the \leroux \wbing method successfully stabilizes the hydrostatic
stratification described in \cref{sec:setup}, even at low resolution.
Furthermore, a series of 2D simulations is presented to estimate the required
strength of the artificial boost of the nuclear energy release.

\subsection{Testing the impact of the \leroux\wbing in 1D and 2D}
\label{sec:testCL}

To demonstrate the capabilities of \wbing, we performed 1D simulations with and
without the \leroux method. For these simulations, the energy input was
switched off.

\begin{figure}
  \centering
  \includegraphics[width=\columnwidth]{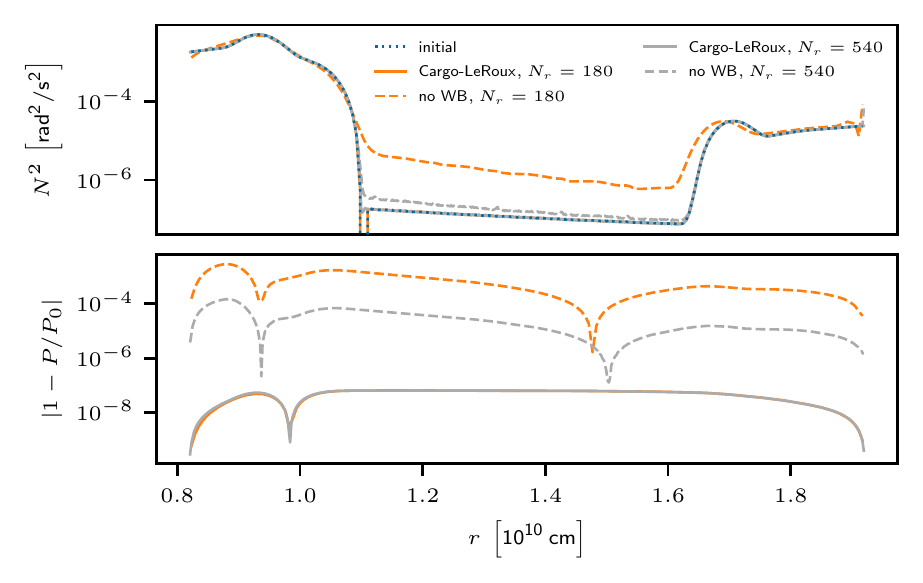}
  \caption{Profiles of the \bvf (\textbf{upper panel}) and relative change in
  pressure (\textbf{lower panel}) after simulating \SI{10}{\hour} of physical
  time with and without \wbing. $N_r$ denotes the number of radial
  cells that are used for the discretization.}
  \label{fig:1D_comparison}
\end{figure}

In \cref{fig:1D_comparison}, the change in the \bvf and the pressure are shown
after simulating \SI{10}{\hour} of physical time (about \num{500} sound
crossing times) in 1D. If the hydrostatic stratification was perfectly
maintained, the initial profiles would stay constant in time. However, for a
grid with 180 radial cells, it is obvious that the formally static setup
changes considerably if no \wbing is applied. The \bvf has increased its value
in the convection zone and the top boundary of the convection zone has moved
inward. The relative pressure change is on the order of \num{e-3} throughout
the domain. In contrast, the \bvf profile does not visibly change in the run
with \leroux\wbing. The relative pressure changes are about \num{e-8} which is
\num{4} orders of magnitude smaller. The simulations shown in
\cref{fig:1D_comparison} apply the lowest radial resolution that is used for
the 3D simulations in \cref{sec:3Dresults}.  The spurious change of the
background stratification is expected to decrease at higher resolutions even if
no \wbing is applied. Indeed, for \num{540} radial cells, the overall changes
decrease considerably. Yet, deviations from the initial stratification are
still visible and the change in pressure is significant.

Hence, the 1D simulations indicate that, especially at low resolution, \wbing
is necessary to maintain hydrostatic equilibrium at a sufficient accuracy. This
is further confirmed in the heated 2D counterparts of the 1D simulations. For a
moderate energy input the setup is evolved for \SI{10}{\hour} of physical time
in 2D wedge geometry.
\begin{figure}
  \centering
  \includegraphics[width=\columnwidth]{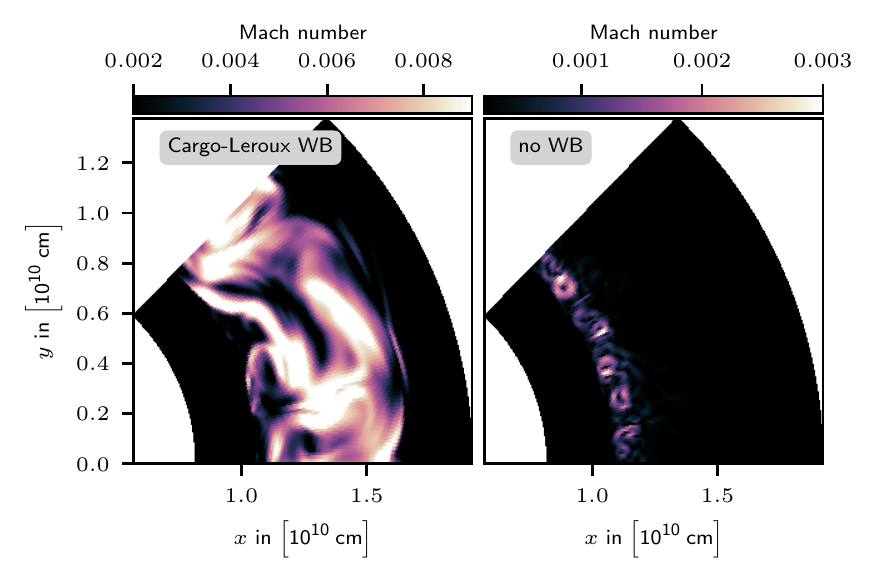}
  \caption{Flow morphology in terms of Mach number in a 2D wedge after simulating
  \SI{10}{\hour} of physical time. The nuclear energy release has been boosted by
  a factor of \num{1e4}. The left panel corresponds to a simulation that applies
  \leroux\wbing while \wbing is absent in the simulation shown on the right. The
  domain is discretized by $\num{180}\times\num{90}$ cells.}
  \label{fig:2D_noCL}
\end{figure}
The resulting flow is depicted in \cref{fig:2D_noCL}. The simulation with
\leroux\wbing has developed the typical large coherent convective eddies inside
the convection zone that are driven by the energy input. This is clearly
different from the flow that develops if no \wbing is applied. Because of the
changing background, the \bvf has become too large, such that the energy input
is not sufficient to establish convection.  Instead, only at the base of the
convection zone where the heating has its maximum a narrow band of small-scale
eddies emerges.

\subsection{Testing artificial boosting of nuclear burning in 2D simulations}
\label{sec:2Dtesting}

Boosting the physical energy generation from nuclear burning is a common
technique in  multidimensional simulations of steady convection, especially in
early stellar evolutionary phases. As predicted by mixing-length theory and
confirmed in numerical studies (see, e.g., \cii or \citealp{andrassy2020a}) the
convective velocity $\vsub{conv}$ scales as
\begin{align}
  \vsub{conv} \propto L^{1/3},
  \label{eq:malum}
\end{align}
where $L$ is the luminosity in the convection zone. Thus, increasing the energy
input leads to larger velocities.

Higher velocities can be beneficial for several reasons. As discussed in
\cref{sec:slh}, convectional  finite-volume schemes based on Riemann solvers
have difficulties to resolve flows at low Mach numbers. Therefore, artificial
boosting can be used to move the flow-velocities to regimes that are more
suitable for the applied numerical scheme. Furthermore, if explicit time
stepping is used, higher velocities improve the ratio of permitted time step
size to the timescale of the flow. This reduces the computational costs.

Another purpose of applying energy boosting is to run simulations with the same
setup but different boosting strengths. This allows one to investigate the
properties of mixing at the boundaries and the entrainment rate as functions of
convective velocities for a single stratification. This has been done in later
phases of stellar evolution for example by \cii or \citet{andrassy2020a} and is
also utilized in \cref{sec:3Dresults}.

The obvious downside of the artificial energy boosting is that the simulations
do not represent the physical situation in the original stellar model anymore.
In 1D stellar evolution codes, the structure of a star critically depends on
the balance between energy generation (e.g., by nuclear burning), cooling
processes (e.g., by escaping neutrinos) and energy transport within the star
(e.g., by radiation or convection). This balance is disturbed if the energy
input is changed. While we think it is still possible with such simulations to
assess the effect of dynamical phenomena such as turbulent mixing and
excitation of waves, they are probably not suitable to study the long-term
behavior of convection where the interplay between turbulence and thermal
diffusion might become important.

Apart from the reasons mentioned above, a sufficient energy boosting is also
necessary to increase the velocity above the numerical threshold at about
$\ma\approx\num{e-3}$ for \slh simulations with the \leroux\wbing method. To
assess by how much the energy generation has to be increased for the 3D
simulations, a set of 2D wedge simulations is performed with varying strengths
of the energy boosting. The resolution is set to $\num{180}\times\num{90}$ cells
(lowest resolution in the 3D runs) and the simulations are performed for boosting
factors ranging from \num{1} (no boosting) to \num{3e4}. The resulting temporal
mean of the root-mean square (\rms) Mach number \mrms as a function of energy input is
then compared to the scaling law in \cref{eq:malum}. In order to determine the
region for which the \rms value of the Mach number is calculated,
the convection zone is identified by means of the gradient of the advected
passive scalar, as will be explained in \cref{sec:pstrace}. For the size of the
time frame, we consider the convective turnover time
\begin{align}
  \si{\tconv} = \frac{2\Delta_\text{CZ}}{\vrms},
  \label{eq:tconv}
\end{align}
where $\Delta_\text{CZ}$ is the radial extent of the convection
zone and $\vrms$ the \rms velocity within the area spanned by
$\Delta_\text{CZ}$ and the horizontal extent of the domain. By taking
\si{\tconv} as the underlying unit of time, we account for the different speed
at which the hydrodynamical processes evolve for different driving strengths.
To finally determine the time frame for which $\vrms$ is determined, we
calculate the number of covered turnover times \Ntconv as
\begin{align}
  \Ntconv(t) = \int_{t_0}^{t} \frac{1}{\tconv(t')}\,\dd t'.
  \label{eq:Ntconv}
\end{align}
Using \cref{eq:Ntconv} automatically accounts for different lengths and
characteristics that may arise for initial transients for different luminosity
boosting and resolutions. Therefore, we find it more convenient to define a
time frame in terms of \Ntconv instead of finding a suitable physical time
interval by hand.

To account for the fact that the boosted region may change during a simulation,
the energy release is integrated over the domain and averaged for the
considered time frame of $t \in \left[t(\Ntconv=10),\, t(\Ntconv=20)\right]$.

\begin{figure}
  \centering
  \includegraphics[width=\columnwidth]{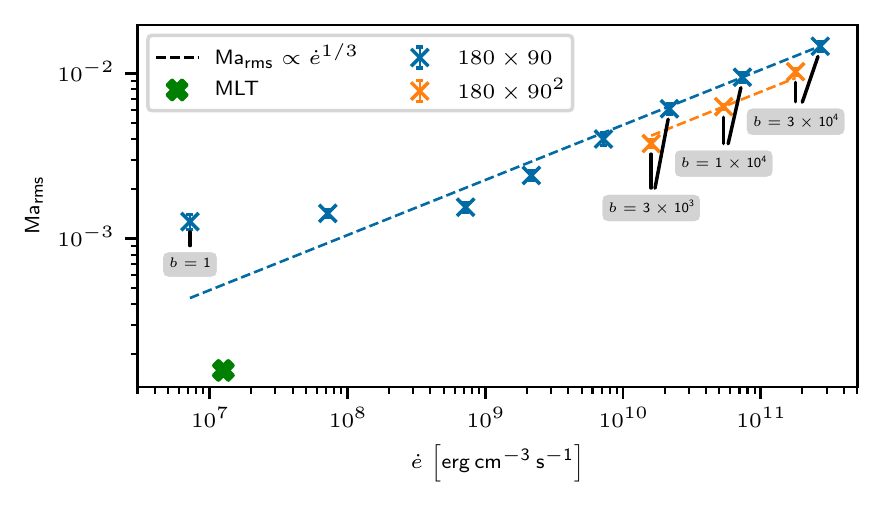}
  \caption{Measured \rms Mach number as function of the input energy rate
  $\dot{e}$ in 2D (blue) and 3D simulations (orange). Vertical error bars
  correspond to the standard deviation of the average over the time frame of
  $\Delta \Ntconv = 10$. The dashed lines reflect the scaling law in
  \cref{eq:malum}. Numbers given in the boxes correspond to energy boosting
  factors for the lowest and the three highest boostings. The green cross
  marks the Mach number of $\ma \approx \num{1.6e-4}$ as predicted by \mlt at
  the nominal energy generation rate in the original \mesa model.}
  \label{fig:2D_3D_rms_scaling}
\end{figure}
As shown in \cref{fig:2D_3D_rms_scaling}, the data points of the three highest
boostings (\numlist{3e3;1e4;3e4}) closely follow the expectation of
\cref{eq:malum}. For lower energy boosting, we find deviations from the
scaling. The corresponding flow patterns along with the detected boundaries are
shown in \cref{fig:mach_pcolor_2D_3D}. The flow of the 2D simulation with the
lowest boosting clearly differs from the other 2D simulations. The appearance
of incoherent, small-scale patterns in \slh simulations of convection is likely
an indication that the flow is driven by numerical artifacts (see also
\citealp{edelmann2021a}). Based on these results, we conclude that the set of
boosting factors $b\in \left[\num{3e3},\num{1e4},\num{3e4}\right]$ is suitable
for the subsequent 3D simulations.

\section{3D \slh results}
\label{sec:3Dresults}

\begin{table*}
\caption{Properties of the 3D simulations with a boosting factor of \num{3e4}.}
\label{tab:res}
\centering
\begin{tabular}{lcccccccc}
\toprule
resolutions & \multicolumn{2}{c}{\lres} & \multicolumn{2}{c}{\mres} & \multicolumn{2}{c}{\hres} & \multicolumn{2}{c}{\vhres} \\
num.\ flux & \ausmpups & \ausmps & \ausmpups & \ausmps & \ausmpups & \ausmps & \ausmpups & \ausmps \bspace\tspace \\
\hline
 $\Delta t_\text{tot}\,\left[\si{\hour}\right]$         & 128.45 & 65.70 & 18.27 & 18.41 & 10.87 & 14.53 & 3.32 & 2.38 \bspace\tspace \\
 $\Delta\text{N}_{\text{conv}}$                         &  44.88 & 21.87 &  6.24 &  5.51 &  3.16 &  4.62 & 1.32 & 0.87 \bspace\tspace \\
 $\overline{\tau}_\text{conv}\,\left[\si{\hour}\right]$ &   3.04 &  2.93 &  2.25 &  2.42 &  2.68 &  2.63 & 2.35 & 2.84 \bspace\tspace \\
 $\mrms\,\left[\num{e-2}\right]$                        &   1.07 &  1.01 &  1.08 &  0.98 &  0.86 &  0.88 & 1.00 & 0.81 \bspace\tspace \\
\hline
\end{tabular}
\tablefoot{
  Simulations with \vhres cells are restarted from the corresponding \hres
  simulations at $\Ntconv =\num{3.1}$ (\ausmpup) and $\Ntconv = \num{3.6}$
  (\ausmp).  Legend: $\Delta t_\text{tot}$: total covered stellar time. $\Delta
  \Ntconv$: total number of turnover times.  $\overline{\tau}_\text{conv}$: mean
  convective turnover time averaged for the last available $0.5\Ntconv$,
  respectively. $\mrms$: \rms Mach number corresponding to
  $\overline{\tau}_\text{conv}$.}
\end{table*}

\begin{table*}
        \caption{Properties of the 3D simulations with a grid size of \lres cells.}
\label{tab:boost}

\centering
\begin{tabular}{lccccccc}
\toprule
boosting & \multicolumn{2}{c}{\num{3e3}} & \multicolumn{2}{c}{\num{1e4}} & \multicolumn{2}{c}{\num{3e4}} \\
num.\ flux & \ausmpups & \ausmps & \ausmpups & \ausmps & \ausmpups & \ausmps \bspace\tspace \\
\hline
 $\Delta t_\text{tot}\,\left[\si{\hour}\right]$         & 199.92 & 183.11 & 139.68 & 97.45 & 128.45 & 65.70 \bspace\tspace \\
 $\Delta\text{N}_{\text{conv}}$                         &  30.61 &  22.61 &  34.00 & 20.70 &  44.88 & 21.87 \bspace\tspace \\
 $\overline{\tau}_\text{conv}\,\left[\si{\hour}\right]$ &   6.13 &   7.24 &   4.27 &  4.61 &   3.04 & 2.93 \bspace\tspace  \\
 $\mrms\,\left[\num{e-2}\right]$                        &   0.40 &   0.33 &   0.65 &  0.56 &   1.07 & 1.01 \bspace\tspace  \\
\hline
\end{tabular}
\tablefoot{Quantities have the same meaning as in \cref{tab:res}.}
\end{table*}

\begin{table*}
\caption{Properties of the 3D simulations with a grid size of \mrest cells.}
\label{tab:boost2}
\centering
\begin{tabular}{lccc}
  \toprule
  boosting & \multicolumn{1}{c}{\num{3e3}} & \multicolumn{1}{c}{\num{1e4}} & \multicolumn{1}{c}{\num{3e4}} \\
  num.\ flux & \ausmpups &\ausmpups & \ausmpups \bspace\tspace \\
  \hline
  $\Delta t_\text{tot}\,\left[\si{\hour}\right]$         & 9.48 & 15.86 & 10.20 \bspace\tspace \\
  $\Delta\text{N}_{\text{conv}}$                         & 1.40 &  4.05 & 4.06 \bspace\tspace  \\
  $\overline{\tau}_\text{conv}\,\left[\si{\hour}\right]$ & 6.86 &  3.85 & 2.39 \bspace\tspace  \\
  $\mrms\,\left[\num{e-2}\right]$                        & 0.33 &  0.61 & 1.09 \bspace\tspace  \\
   \hline
\end{tabular}
\tablefoot{Quantities have the same meaning as in \cref{tab:res}.}
\end{table*}

After the basic properties of the stellar model have been tested in 1D and 2D
hydrodynamic simulations, this section presents the results regarding turbulent
flow properties and entrainment obtained from 3D simulations. We analyze the
results for varying resolution and convective driving. To demonstrate that the
low-Mach \ausmpup flux scheme is beneficial even for moderate Mach numbers, the
respective results are compared to its basic version \ausmp that is not
expected to show enhanced low-Mach capabilities. A comparison to more commonly
used flux functions, such as the \roe solver, would have been a more obvious
choice. This was not possible as the applied \leroux\wbing method is not fully
compatible with the \roe scheme.  However, in \cref{sec:spectra}, we show for a
reduced domain that the numerical diffusivity is similar for \ausmp and the
\roe scheme.

A major restriction for our 3D simulations is posed by the available
computational resources. While a higher resolution is certainly desirable, it
considerably reduces the physical time for which we could follow convection.
However, convection has to be covered for several turnover times \si{\tconv} in
order to analyze mixing processes at the boundaries of the convection zone. We
therefore can only investigate the effect of boundary mixing at the lowest
resolution of \lres. At higher resolution, our simulations only cover a few
multiples of \tconv which is too short to track mixing at the boundaries but is
sufficient to extract properties of turbulence. In this section we present
simulations with resolutions ranging from \lres to \vhres. The basic properties
of the simulations are summarized in \cref{tab:res,tab:boost,tab:boost2}. We
note that the radial resolution from \mres to \hres cells changes by a factor
of \num{1.5}, while the corresponding number of horizontal cells
changes by a factor of \num{2}. This was done inadvertently, but we are
confident that it does not prevent the comparison of the results between the
different resolutions. The simulations at a resolution of \vhres cells are
restarted from the corresponding simulations at \hres at a stage of fully
developed convection. This avoids the slow initial transients and hence reduces
computational costs.

To conclude the 2D scaling test of the previous section, the scaling relation
\cref{eq:malum} is shown for the lowest resolution and the \ausmpup solver in
\cref{fig:2D_3D_rms_scaling} (orange crosses). The 3D data is in good agreement
with the expected scaling. The corresponding flow patterns are found in
\cref{fig:mach_pcolor_2D_3D} at a resolution of \lres. From \mlt, a convective
velocity of $\ma_\text{MLT}\approx\num{1.6e-4}$ is predicted. If we extrapolate
from the 3D results to stellar luminosity we find a value of
$\ma_\text{ext}\approx\num{4.0e-4}$. The ratio $\ma_\text{ext} /
\ma_\text{\mlt} \approx 2.5$ is similar to what has been obtained by
\citet{jones2017a}. This ratio indicates a reasonable agreement, taking into
account that \mlt only provides an order-of-magnitude estimate and that the
results from our simulations need to be extrapolated to nominal luminosity.

\subsection{Tracing the boundaries of the convection zone}
\label{sec:pstrace}

For the analysis that is presented in the subsequent sections, the top and
bottom radii of the convection zone, \rczb and \rczt, have to be extracted from
the simulations. This can be done in different ways, for example by considering
the radii where the decline in the horizontal velocity is steepest
\citep{jones2017a}, where the species abundance gradient is steepest
\citep{meakin2007b}, or where the mean atomic weight is equal to its averaged
value within the convective and adjacent stable zone (\ci).  Furthermore, to
avoid the use of averages which might underestimate the effect of strong but
rare mixing events, extreme value statistics could be used \citep{pratt2017a}
or the standard deviation of a dynamical quantity like the kinetic energy
\citep{higl2021a}.

\begin{figure}
  \centering
  \includegraphics[width=\columnwidth]{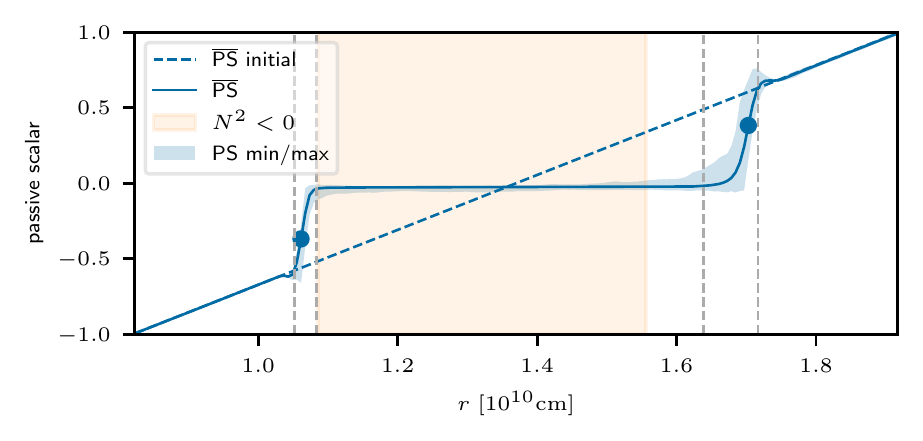}
  \caption{Horizontal mean of the advected passive scalar for a simulation with
  a resolution of \lres. The initial distribution of the scalar is shown as
  dashed line, the solid line represents the time-averaged profile for $t
  \in \left[\tofn{9.9}, \tofn{10.1}\right]$. The blue shaded area corresponds
  to the minimal and maximal value of the passive scalar at the corresponding
  radius for the latest considered snapshot. The radii at which the absolute
  value of the radial derivative is largest are indicated by dots. They
  define the position of the top and bottom boundaries. The shaded orange area
  marks the convective region according to the stability criterion $N^2<0$.
  Vertical dashed lines denote the respective boundary widths which are defined
  in \cref{sec:bwidth}.}
  \label{fig:line_cb_tracing}
\end{figure}

Our approach is to advect $\rho X$ with the fluid flow where $X$ is a passive
scalar. Its initial distribution increases at a constant slope from \num{-1} at
the bottom boundary of the computational domain to \num{1} at the top (dashed
line in \cref{fig:line_cb_tracing}). The passive scalar will quickly be mixed
within the convection zone while forming a transition between the initially
linear decrease and a flat region. The position of the boundaries is then
defined as the radius where the spatial gradient of the horizontally averaged
passive scalar is largest. We find that after an initial redistribution of the
scalar by the onset of convection our method gives robust results.

This definition of the boundary position is similar to using the abundance
gradient. However, it does not depend on the initial 1D structure in terms of
strength and position of the gradients. Furthermore, the abundance and passive
scalar profiles are immediate measurements of the mixing compared to measuring
for example overshooting events or standard deviations, which are linked to
mixing only indirectly.

For an exemplary simulation, \cref{fig:line_cb_tracing} shows the profile of
the advected passive scalar at the start of the simulation and around
\tofn{10}. The efficient mixing by convection has homogenized the passive
scalar within the convection zone. At the top and bottom boundary of the
computational domain, the profile of the passive scalar is almost
not distinguishable from the initial distribution. The orange shaded area denotes
the convection zone according to the criterion $N^2<0$ and it is clearly
visible that this definition underestimates the extent of the convection zone.
The bottom and top boundaries of the convection zone as given by the maximum
absolute value of the radial gradient of the passive scalar are shown as blue
dots. There are small amounts of numerical under- and overshoots in the
profile of the passive scalar. This is due to a lack of slope-limiter for
reconstruction for the implicit time stepping.

\subsection{Kinetic energy spectra and comparison with turbulence theory}
\label{sec:spectra}

\citet{kolmogorov1941a} (see also \citealp{landau1987a}) predicts that the
spectrum of kinetic energy $\varepsilon_\mathrm{kin}$ in 3D isotropic
turbulence follows
\begin{align}
  \varepsilon_{\mathrm{kin}}(\ell) \propto \hat v^2(\ell)
  \propto \ell^{-5/3},
  \label{eq:kolmo-sc}
\end{align}
where $\ell$ is the angular order. Although stellar convection is not isotropic
on large scales, many numerical experiments reveal spectra similar to this
prediction on sufficiently small scales
(\citealp{porter2000a,gilet2013a,verma2017a}, \ci). The spectra of turbulent
convection in 3D typically divide into three regions (see \citealp{arnett2015a}
for a more detailed discussion): At large spatial scales, that is at small
values of the angular order $\ell$, the energy from heating is injected into
the flow, forming the integral range. At somewhat smaller scales, or
equivalently for larger $\ell$, the inertial range forms that follows the
scaling of \cref{eq:kolmo-sc}. The inertial range extends down to the small
scales where dissipating effects such as viscosity become relevant and
turbulent kinetic energy is transformed into internal energy. This leads to a
steeper drop in $\varepsilon_{\mathrm{kin}}(\ell)$ for larger $\ell$ and marks
the dissipation range. In the stellar context, it is impossible to resolve the
spatial scales where physical viscosity takes place. Thus, in implicit large
eddy simulations (\iles), the effect of viscosity is not modeled explicitly but
follows from the numerical viscosity inherent in the applied numerical scheme
at small scales \citep[see, e.g.,][]{arnett2015a,arnett2018a}. This is the case
for the \slh code that solves the Euler equations which follow from the
Navier-Stokes equations for vanishing viscosity but does not apply any subgrid
scale model for turbulent dissipation. It therefore is desirable to resolve the
scaling \cref{eq:kolmo-sc} to the smallest scales possible while still having a
numerically stable scheme.  Hence, one way to compare the quality of numerical
schemes is to compare their respective range in $\ell$ for which they reproduce
an inertial range with a scaling according to \cref{eq:kolmo-sc}.

In the following, we present the spectra for the 3D \slh simulations and
compare the low-Mach \ausmpup solver to the \ausmp scheme. In setups with an
approximate spherical symmetry, the spatial spectra of turbulent flows are
typically given in terms of power spectra for spherical harmonics. This makes
use of the fact that a given function $F(\vartheta, \varphi)$ on the spherical
surface can be decomposed into spherical harmonics as
\begin{align}
  F(\vartheta, \varphi) = \sum_{\ell=0}^{\infty}\sum_{m=-\ell}^{\ell} f_{\ell m} Y_{\ell m}(\vartheta, \varphi),
  \label{eq:sh}
\end{align}
where $f_{\ell m}$ is the amplitude for the corresponding spherical harmonic
$Y_{lm}$ of angular degree $\ell$ and angular order $m$. For our analysis we
apply the open-source
\sht\footnote{\href{https://shtools.oca.eu}{https://shtools.oca.eu}}
\citep{shtools}, a collection of Fortran90 and Python libraries for spherical
harmonics data analysis. To decompose the velocity fields that result from our
simulations, we proceed as follows:

\begin{figure}
  \centering
  \includegraphics[width=\columnwidth]{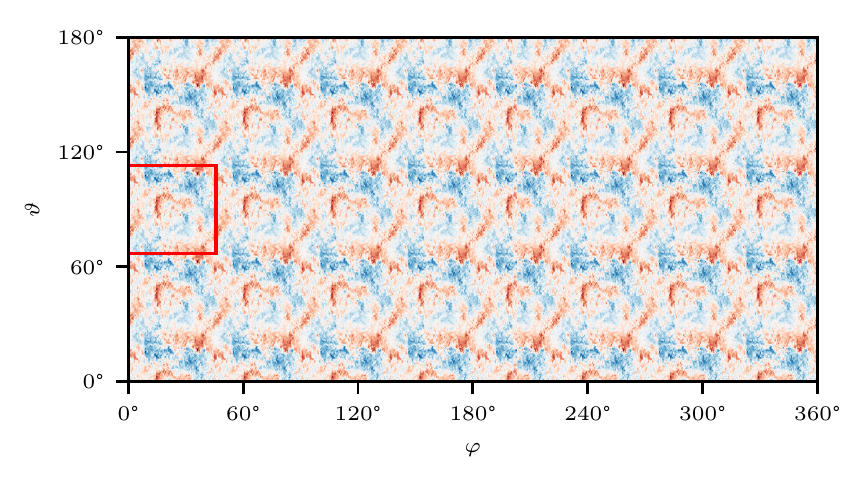}
  \caption{Expansion of the velocity data for one exemplary 3D wedge
  simulation. Color coded is the velocity component in $\varphi$-direction.
  The red square marks the actual computational domain. The rest of
  the plane is filled by periodically repeating the simulation data in
  $\vartheta$ and $\varphi$ direction.}
  \label{fig:spec3Dpad}
\end{figure}
The \sht assume that the input data is provided for the whole spherical
surface. The computational domain in our simulations, however, is a spherical
wedge. We therefore expand the $\varphi-\vartheta$ plane covered by our
simulations to the full spherical surface. For this, the data from our
simulation is repeated periodically to fill the regions that are not covered by
the computational domain, see \cref{fig:spec3Dpad}. This gives slightly weaker
artifacts than zero-padding.

To further reduce the artifacts introduced by our limited domain, we make use
of the ability of \sht to apply an arbitrary window function to extract
localized spectra. The \sht construct the windows automatically and provide the
user the option to restrict the bandwidth of the created windows by an upper
limit $\ell_\text{max}$. The necessary bandwidth of the window function
increases with smaller localized areas. The \sht then calculate different
realizations of window functions that have their power concentrated in the
considered region within the $\vartheta-\varphi$ plane. The spectra of all
windows are averaged (multitaper, see also \citealp{wiezorek2005a}). For the
input of the multitaper spectrum, we set $\ell_\text{max}$ to a sufficiently
high value to obtain at least \num{10} window realizations from \sht which have
\SI{99}{\percent} of their power localized in the computational domain.

The spectra that are presented in the following are taken at a
constant radius in the middle of the convection zone. This is justified if the
flow is isotropic, which is not the case at large scales (small $\ell$) but a
reasonable assumption at small scales. Isotropy is also a necessary condition
for the Kolmogorov-scaling \cref{eq:kolmo-sc} to form. The extracted spectra
are averaged over roughly one convective turnover time.

\begin{figure}
  \centering
  \includegraphics{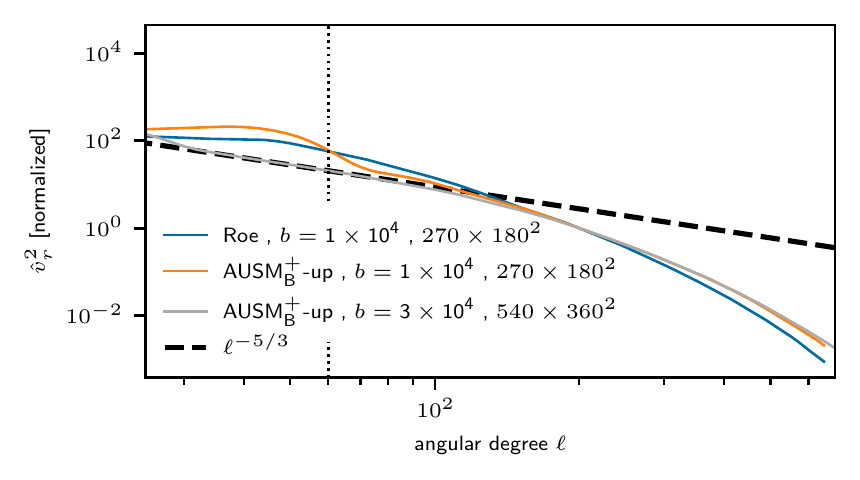}
  \caption{Spectra of the radial kinetic energy component. The blue and orange
  line correspond to simulations with the \roe and \ausmp schemes of a reduced
  domain that only contains a fraction of the convection zone. The gray line
  shows the spectrum for a simulation of the full domain with the \ausmp
  solver, the same grid spacing, but a different energy boosting. The
  amplitudes have been normalized such that they are unity at $\ell=200$ to
  ease the comparison. The dashed line marks the Kolmogorov-scaling according
  to \cref{eq:kolmo-sc}. The vertical dotted line at $\ell_\text{max}=60$
  denotes the spectral width of the applied window functions for the runs with
  $270\times180^2$ cells. For $\ell\leq \ell_\text{max}$, their spectra are
  dominated by the convolution with the window function and do not reflect
  real data. The horizontal axis is truncated at the spectral width of the
  window function for the \hres run which corresponds to $\ell_\text{max}=25$.}
  \label{fig:spec_roevsausm}
\end{figure}

In order to demonstrate the improved performance of the low-Mach \ausmpup
solver, we compare it to its basic variant \ausmp. The AUSM solver family is
not yet widely used in the astrophysical community. To get an idea how
\ausmp compares to the well-known \roe scheme, we compare their spectra for a
reduced domain, which only contains a subset of the convection zone. This is
necessary because we find numerical artifacts for \roe in combination with \leroux
\wbing that lead to spurious flows in the stable regions at radii where
abundances change.

Their spatial resolutions are the same as for the \hres simulations of the full
domain. The result is shown in \cref{fig:spec_roevsausm}. The spectra
demonstrate that, at least within the convection zone, the \ausmp and the \roe
solver are both dissipative and do not show an inertial range. For comparison,
the spectrum of a simulation which has the same spatial resolution but using
the \ausmp scheme is added to the figure along with the Kolmogorov-law
\cref{eq:kolmo-sc}. The similarity between the \roe and \ausmp solver is also
evident in the flow pattern, \cref{fig:mach_pcolor_roe_ausm}.

\begin{figure}
  \centering
  \includegraphics{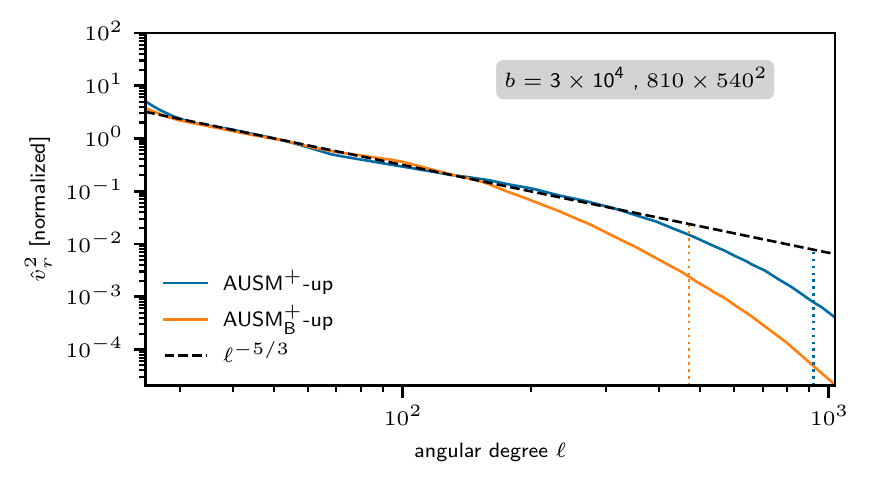}
  \caption{Spectra of the radial kinetic energy component for the \ausmp and
  \ausmpup solver at a resolution of \vhres cells and an energy boosting of
  $b=\num{3e4}$. The amplitudes are normalized to unity at $\ell=50$ for better
  comparability. Dotted vertical lines mark the angular degree $\ell$ at which
  the relative deviation from the Kolmogorov-law is one decade. For the \ausmp
  solver this happens at $\ell \approx 470$, for the \ausmpup solver at $\ell
  \approx 920$.  The horizontal axis is truncated at the spectral width of the
  applied window function ($\ell_\text{max}=25$).}
  \label{fig:spechres}
\end{figure}
The spectra for the highest available resolution and the full domain are shown
in \cref{fig:spechres}. We find that the \ausmp and \ausmpup solver both show a
well-defined inertial range where the slope closely follows the prediction of
a Kolmogorov spectrum. The vertical dotted lines in \cref{fig:spechres} mark
the scale at which there is a significant deviation from the Kolmogorov-law.
From this measure we find that the inertial range of the \ausmpup solver
extends toward scales that are about a factor two smaller compared to the
\ausmp solver.

\begin{figure*}
  \centering
  \includegraphics{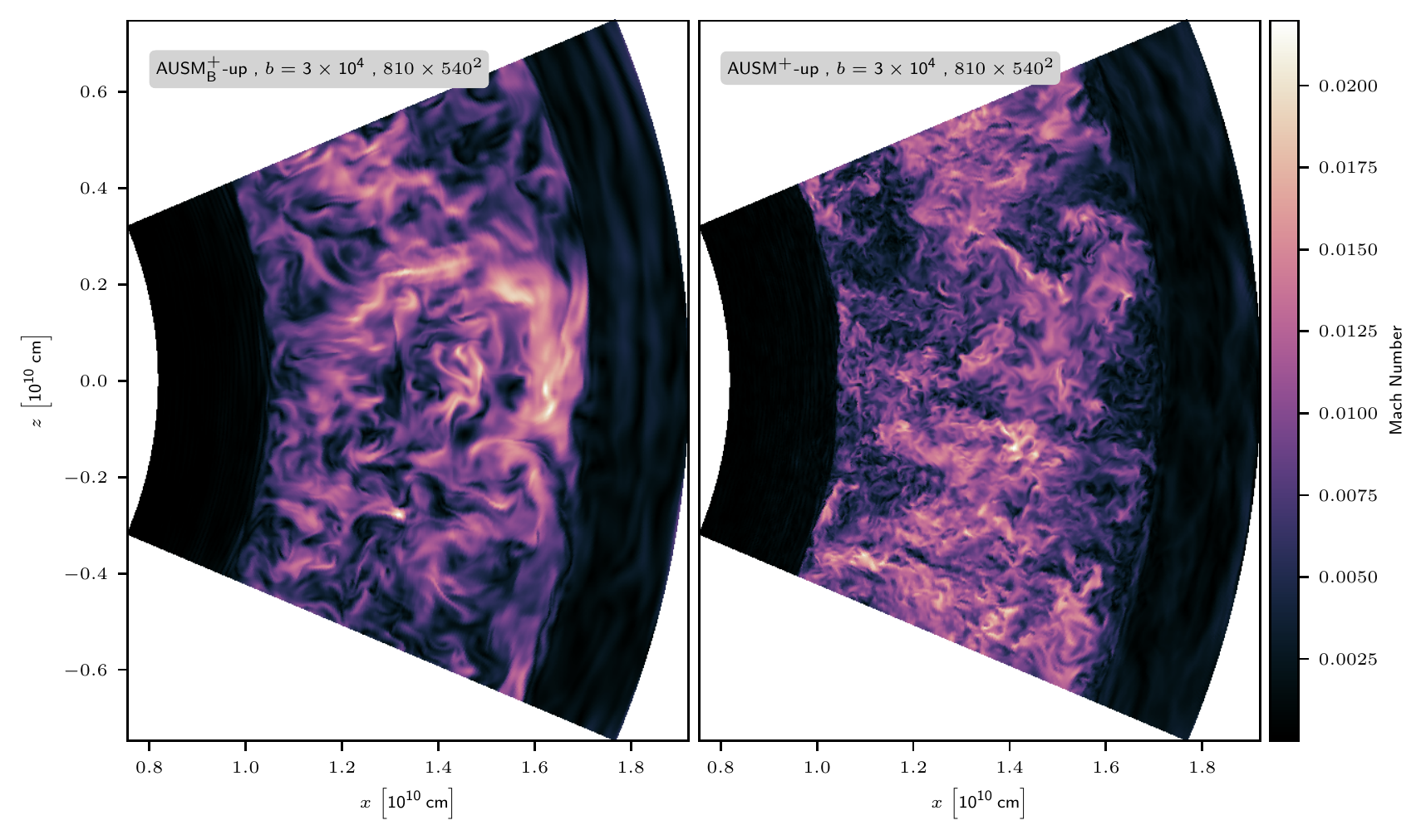}
  \caption{Mach number for a slice through the domain for simulations at a
  resolution of \vhres cells with the \ausmp solver (left) and \ausmpup solver
  (right). The data is taken from the latest available snapshot, respectively.}
  \label{fig:pcolor_flopat}
\end{figure*}
\begin{figure*}
  \centering
  \includegraphics[width=\textwidth]{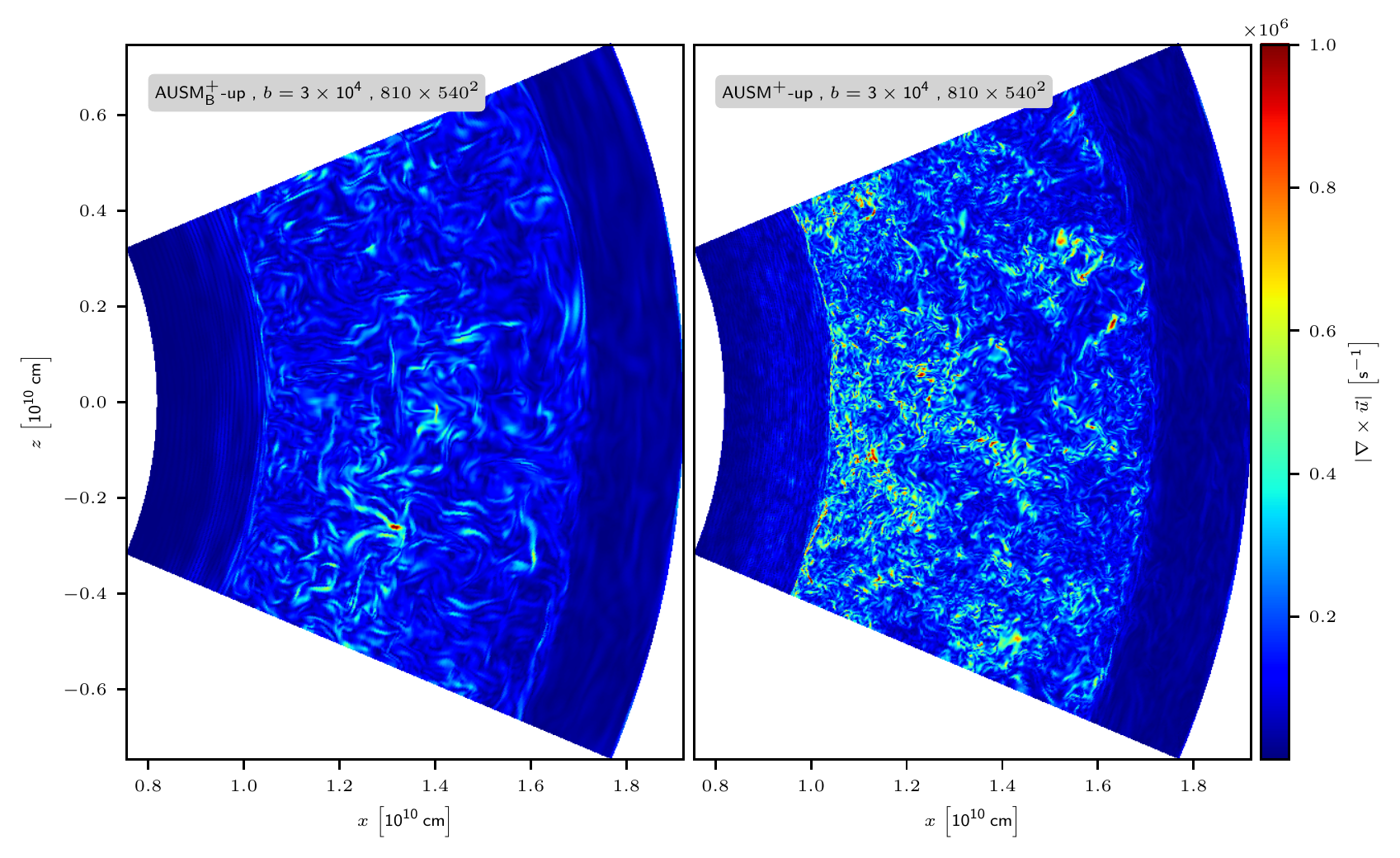}
  \caption{Similar to \cref{fig:pcolor_flopat}, but for the magnitude of the
  vorticity.}
  \label{fig:pcolor_flopat_vort}
\end{figure*}
In \cref{fig:pcolor_flopat} the turbulent convective velocity field is depicted
for a slice through the three-dimensional domain for a single snapshot. The
\ausmpup scheme clearly shows smaller structures in the flow field as compared
with \ausmp on the same computational grid. This is also apparent in
\cref{fig:pcolor_flopat_vort} which shows the magnitude of vorticity
$\left|\nabla \times \vec u\right|$, where $\vec u$ is the velocity vector.
\begin{figure}
  \centering
  \includegraphics{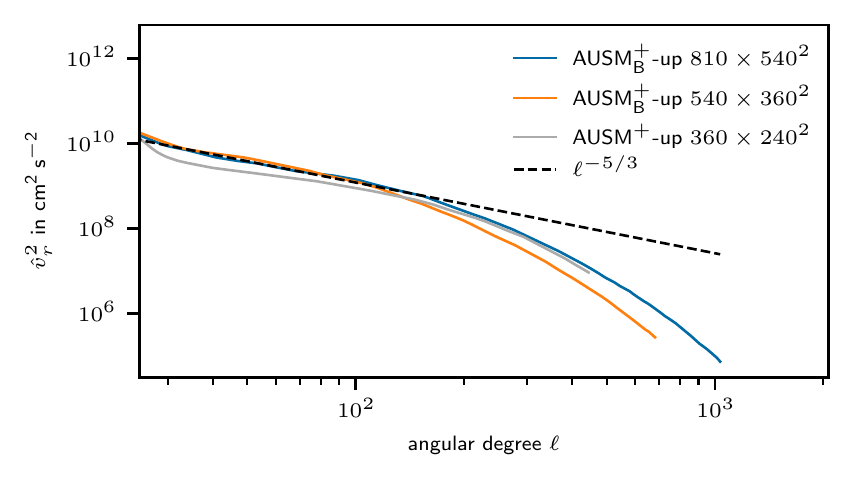}
  \caption{Spectra of the radial kinetic energy component for the \ausmp and
  \ausmpup solver at different resolutions and an energy boosting of
  $b=\num{3e4}$.}
  \label{fig:specvs}
\end{figure}
To further illustrate the advantages of the low-Mach flux \ausmpup over \ausmp,
we compare in \cref{fig:specvs} the spectra at different resolutions. We find
that for the \ausmpup solver, a grid resolution of \mres gives an inertial range
that is comparable to the \vhres resolution of simulations with \ausmp.

\subsection{Comparing numerical dissipation from \rans results}
\label{sec:rans}

The comparison of the kinetic energy spectra is complemented by analyzing the
different simulations in the terms of Reynolds averaged implicit large eddy
simulations (\rans) \citep{mocak2014a,arnett2019a}. The fundamental idea is to
separate the different components of the Navier-Stokes equations into mean and
fluctuating parts and to determine them by analyzing numerical simulations. The
physical interpretation of these parts then gives useful insight into the complex
interplay between different processes that act in turbulent convection and at
the boundaries of the convection zone.

While the \rans framework provides a wealth of equations (see
\citealp{mocak2014a} for an extensive overview), we focus on analyzing the
equation for turbulent kinetic energy. It allows one to quantify the effect of
implicit numerical dissipation of kinetic energy that is inherent in all
\iles. This equation has been used in several publications in the past
\citep[see, e.g.,][]{arnett2009a,mocak2014a,mocak2018a}, and aided the analysis
of the effects of resolution and convective driving (\ci, \cii) or different
strengths of stratification \citep{viallet2013b}. Following the formulation of
\citet{mocak2014a}, the time evolution for the kinetic energy of an inviscid
fluid can be written as
\begin{align}
  \partial_t(\bar\rho\tilde{\epsilon}_k) +\nabla_r (\bar\rho\,\tilde{v}_r\tilde\epsilon_k) &=
    - \nabla_r\left(f_P + f_k\right) + W_b + W_P,
\label{eq:rans_ekin}
\end{align}
where
$\epsilon_k$ is the specific kinetic energy,
$f_P = \overline{P'v_r'}$ the acoustic flux,
$f_k = \overline{\rho v_r'' \epsilon_k''}$ the turbulent kinetic energy flux,
$W_b = \overline{\rho}\,\overline{v_r''}\,\tilde{g}_r$ the buoyancy work,
$W_P = \overline{P'd''}$ the turbulent pressure dilatation, and
$d = \nabla\cdot \vec v$.
The radial component of the gravitational acceleration is denoted by $g_r$. The
definition of the Reynolds average $\overline{q}$, Favre average $\tilde{q}$,
and the corresponding fluctuations $q',\, q''$ for a quantity $q$ are given in
\cref{appendix:reynolds}. For a more detailed discussion of the individual
terms, see for example \citet{meakin2007b}, \citet{viallet2013b}, or
\citet{mocak2014a}.

Because numerical solutions are only approximations to the true solution,
\cref{eq:rans_ekin} will in general not be fulfilled exactly in hydrodynamic
simulations. Instead, there will be a residual \nekin between the left-hand and
right-hand side. In energy conserving methods like finite volume schemes,
\nekin then measures the numerical dissipation of kinetic energy into internal
energy, the fundamental property of \iles. The exact value of \nekin depends on
the details of the numerical scheme, the resolution, but also on the specific
physical problem at hand. Generally, the value of \nekin cannot be
controlled in \iles.  However, extracting the terms in \cref{eq:rans_ekin} from
a hydrodynamic simulation, the strength of numerical dissipation that acted
for the considered time in a specific simulation can be determined from the
average value of \nekin.

\begin{figure*}
  \centering
  \includegraphics[width=\textwidth]{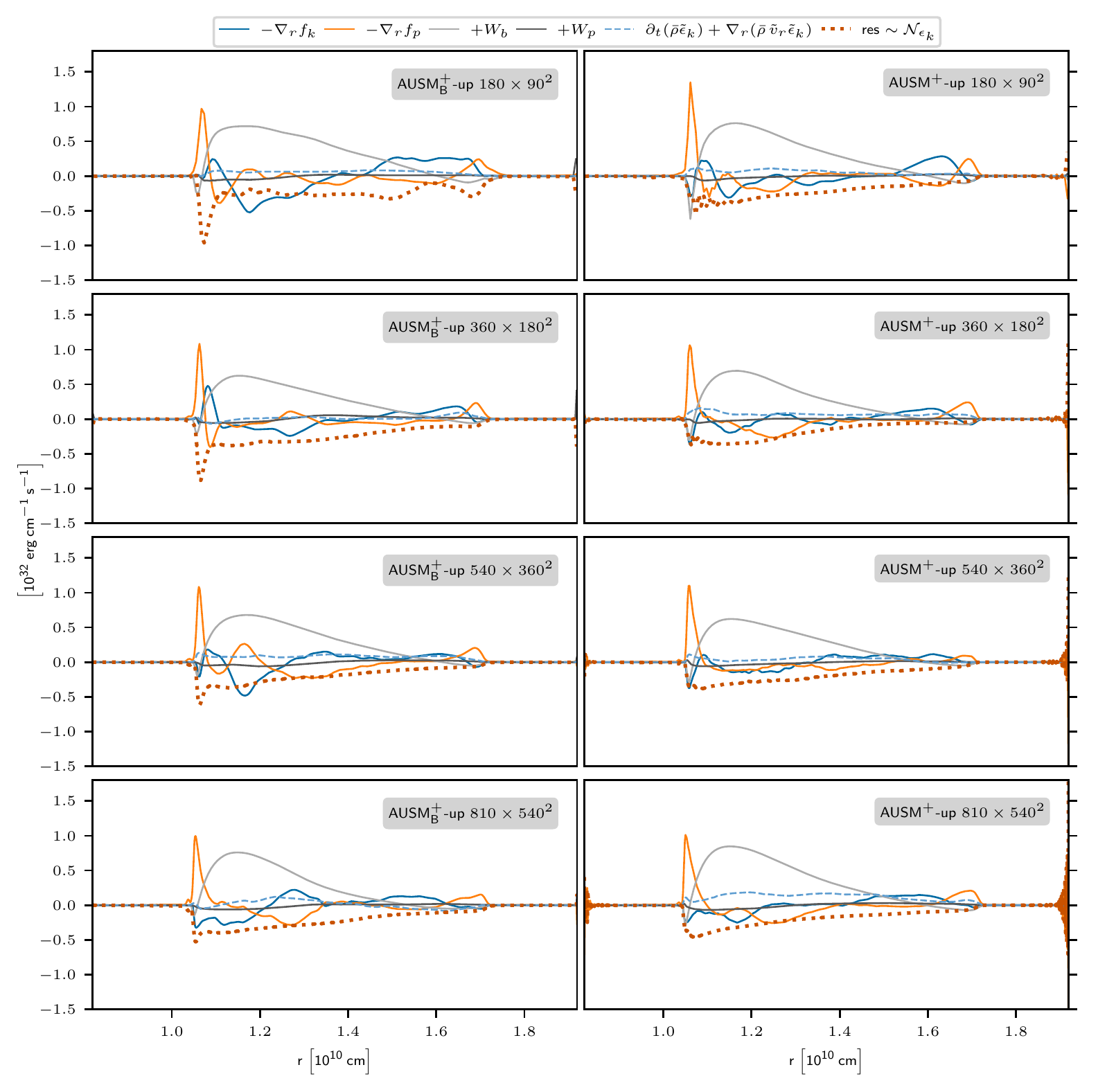}
  \caption{Profiles of the different components of kinetic energy equation in
  the \rans framework, \cref{eq:rans_ekin}. The resolution increases from top
  to bottom. The left column corresponds to results using the \ausmp solver and
  the right column to results using the \ausmpup solver. All simulations boost
  the nuclear energy generation by a factor of \num{3e4}. The profiles are
  averaged for roughly one convective turnover time. Similar to
  \citet{viallet2013b}, the components are multiplied by the geometrical factor
  $4\pi r^2$.}
  \label{fig:line_rans_b3e4}
\end{figure*}
We calculate all the terms in \cref{eq:rans_ekin} for the \ausmp and \ausmpup
solver at different resolutions. Third-order central differences are used to
evaluate the gradients. Except for the highest resolution, the results are
averaged over the time interval of $t\in\left[\tofn{2},\tofn{3}\right]$  which
is the maximum overlapping time frame. For the highest resolution, the
simulations are averaged over only $\Delta\Ntconv = 0.6$.  Ideally, the
averages would be performed over several turnover times to improve the
statistics. While our short time frames probably make a quantitative comparison
of the components less robust, we think that a qualitative comparison is still
meaningful and that the main characteristics of \cref{eq:rans_ekin} are
captured.

In \cref{fig:line_rans_b3e4} the profiles of the individual terms of
\cref{eq:rans_ekin} are depicted for successively increasing
resolutions\footnote{In the \rans analysis framework of \slh, all required
fluctuations are calculated and stored already during the simulation, such that
we have data for every single time step. However, there was a flaw in the
calculation of the velocity divergence for the simulations presented here.
Therefore, the velocity divergence had to be re-calculated in a post-processing
step, for which the 3D velocity data is only available for the stored grid
files but not for every time step. Fortunately, only the value of $W_P$ is
affected which is, however,  small in general and the impact of the
post-processing is negligible.}. We find our results in qualitative agreement
with simulations of oxygen burning \citep[][\figr 8]{viallet2013b} and carbon
burning (\figr{9} of \ci, see also \cii). The small but noticeable
nonzero values for the left-hand side of \cref{eq:rans_ekin} (dashed lines in
\cref{fig:line_rans_b3e4}) are due to the short time interval considered.
\begin{figure*}
  \centering
  \includegraphics{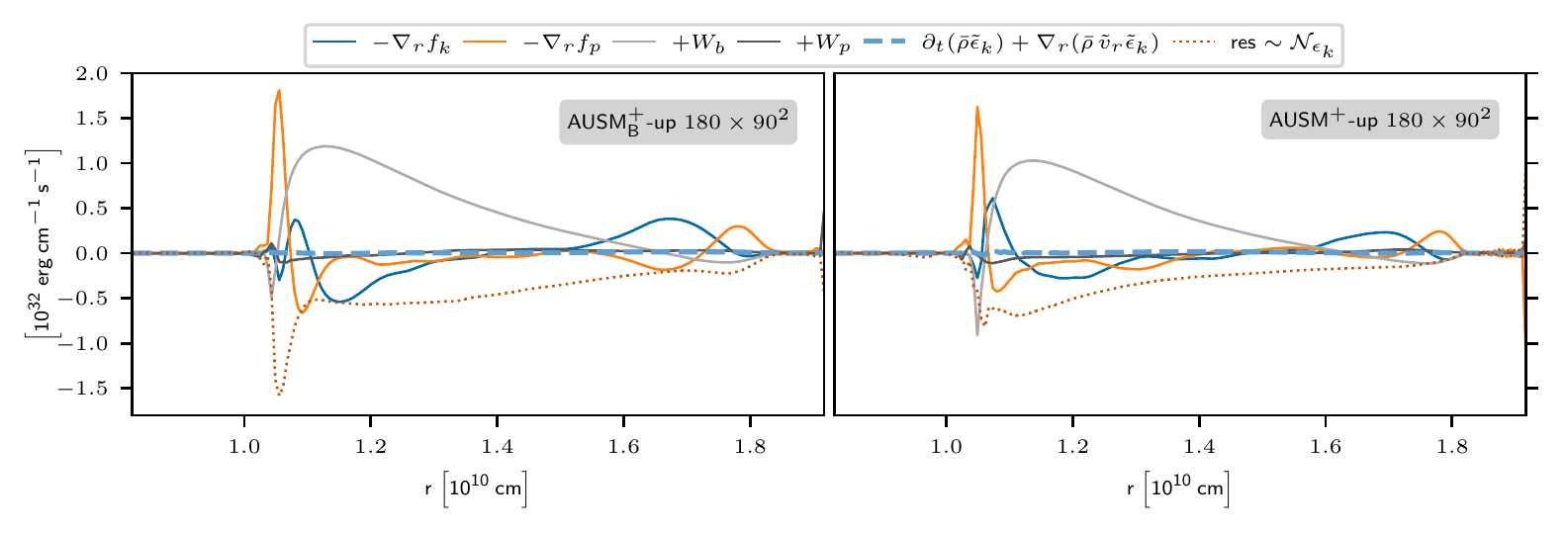}
  \caption{Same quantities as in \cref{fig:line_rans_b3e4}, but for a time interval of
  $t\in\left[\tofn{10},\tofn{20}\right]$.}
  \label{fig:line_rans_b3e4_180}
\end{figure*}
For
comparison, we recalculated in \cref{fig:line_rans_b3e4_180} the results for
the lowest resolution but a time interval that covers $\Delta\Ntconv = 10$.
Here, the time evolution of the kinetic energy is close to zero.

The dominant part on the right-hand side is the buoyancy work $W_b$ that is
positive in the convection zone and changes sign at the boundaries to the
stable layers. The acoustic and turbulent kinetic energy fluxes show a more
complex behavior and change signs several times in the convection zone. The
pressure dilatation term $W_P$ takes a rather low value for all simulations
owing to the fact that the density stratification within the convection zone is
small. As shown by \citet{viallet2013b}, the situation can be different in
other setups. They find that in the convective envelope of a \SI{5}{\msol} red
giant star, pressure dilatation contributes a significant part to the overall
budget of \cref{eq:rans_ekin} as the convection zone spans several pressure
scale-heights. In general, we do not find significant qualitative differences
between different resolutions and between the \ausmp and \ausmpup solver. At
the lowest resolution with \ausmpup, small oscillations on the grid level are
visible for the acoustic flux $f_P$ within the convection zone. However, they
vanish for increasing resolution. At high resolution, we find oscillations in
$f_P$ at the domain boundaries, the origin of which is not completely clear. We
suspect an interplay of better resolved internal gravity waves with the
constant ghost cell boundaries. This inevitably leads to shear because
velocities are set to zero in the boundary cells.

The dotted lines in \cref{fig:line_rans_b3e4} correspond to the numerical
dissipation of kinetic energy \nekin acting in the respective simulation. In
general, the dissipation is distributed over the whole convection zone and
vanishes in the stable layers. For the \ausmpup solver, some smaller
oscillations are visible near the boundaries which stem from the oscillations
in $f_P$. Comparing the results of the \ausmp and \ausmpup solver at each
resolution, the profiles of \nekin have similar amplitudes in the main part of
the convection zone. However, for the simulations using the \ausmp solver, the
numerical dissipation shows a distinct peak at the bottom boundary. The peak
height and width decreases with increasing resolution. The same behavior was
found by \ci for carbon-shell burning and by \citet{viallet2013b} for
oxygen-shell burning. However, this peak is absent in the simulations using the
\ausmpup solver. From the shape and position of the peak of \nekin in the plots
for the \ausmp solver it seems that the peak is due to an imbalance between
the acoustic flux $f_p$ and $W_b$. Although the peak in $f_p$ appears to be
similar in shape and amplitude for the two solvers, a more pronounced opposed
peak in $W_b$ seems to counteract the gradient of $f_p$ in the \ausmpup runs.

\begin{figure}
  \centering
  \includegraphics[width=\columnwidth]{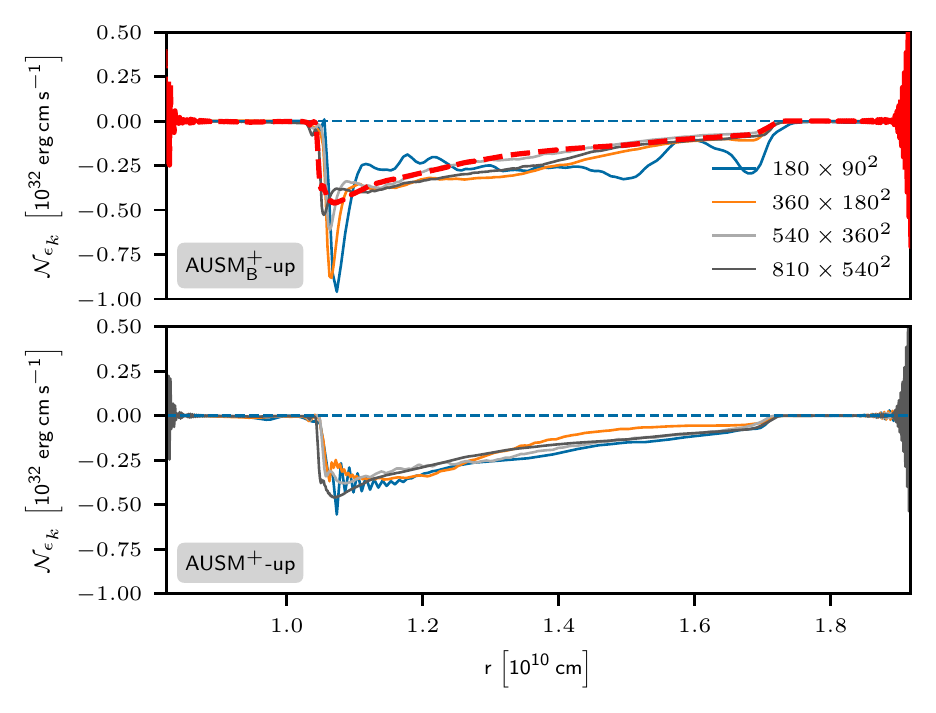}
  \caption{Residual \nekin for the runs shown in \cref{fig:line_rans_b3e4}. For
  comparison, the profile of \nekin for the \vhres \ausmpup simulation is
  included as a red dashed line in the upper panel.}
  \label{fig:line_rans_residual}
\end{figure}
We directly compare the numerical dissipation \nekin for the different
simulations in \cref{fig:line_rans_residual}. For \ausmpup, the amplitudes seem
to be converged already for the lowest resolution, although low-resolution runs
show oscillations in the numerical dissipation. Also for the \ausmp solver, the
dissipation in the bulk of the convection zone seems not to depend strongly on
the resolution. This is consistent with the expected independence of the
turbulent dissipation rate from the effective viscosity, which is set by the
grid scale.  However, at the bottom boundary, the peak decreases with
increasing resolution and seems to converge toward a value that is similar to
the dissipation of the \ausmpup solver. These results are fully in line with
the  simulations shown in \figrs{1 and 2} of \citet{arnett2018a} which
summarize the numerical dissipation in oxygen- and carbon-shell burning
simulations with the PROMPI code. For their highest-resolution run, they find
that the peak at the bottom boundary seems to merge with the bulk dissipation.
This indicates that the low-Mach \ausmpup solver improves the behavior at the
bottom boundary, even at moderate Mach numbers and moderate resolution.

\subsection{Convective boundary mixing}
\label{sec:cbm}

An important process in stellar interiors is the entrainment of material from
stable layers at the boundaries of a convection zone into the convective
region. This has implications for the star's further evolution because the
entrained material serves as fuel for the burning region. Despite its
importance, it needs to be parametrized in conventional 1D stellar evolution
simulations due to its inherently multidimensional nature. For the various types of
convection, for example in shallow zones in the stellar interior or extended
regions in the envelope of solar-like stars, different physical mechanisms are
dominant. It is therefore of interest to develop and validate different
possible parametrizations with the help of multidimensional simulations.

\citet{viallet2015a} suggest to use the local \pec number, the ratio of
advective and diffusive timescales, in the boundary region to distinguish
between different types of convective boundary mixing. Estimating the typical
velocity $v$ and length scale $l$ of convection through \mlt, we find a minimum
\pec number within the convection zone of
\begin{align}
  \text{Pe} = \frac{ul}{K} = \frac{3D_\text{mlt}}{K} \approx \num{5e4}\gg 1,
\end{align}
where $K$ is the thermal diffusivity [see \cref{eq:tdiff}] and
$D_\text{\mlt}={1}/{3}\,u_\text{\mlt}\,l_\text{\mlt}$ is the diffusion coefficient
obtained from \mlt. The large \pec number implies minor importance of radiation for
the mixing, in accordance with our estimates in \cref{sec:thermaldiff}. The
artificial boosting will increase the \pec number even further in the
hydrodynamic simulations. Following \citet{viallet2015a}, at $\text{Pe}\gg 1$
mixing can be thought to occur via turbulent entrainment, where small-scale
instabilities are caused by the shear created by overturning convective
cells at the boundaries (see \citealp{viallet2013a} for a detailed
description). As demonstrated by \citet{meakin2007b} for stellar convection,
turbulent entrainment can be characterized in terms of the bulk Richardson
entrainment law
\begin{align}
  \frac{\ve}{\vrms} = A\, \rib^{-n},
  \label{eq:bri-scaling}
\end{align}
where \ve is the entrainment velocity of the top or bottom convective
boundary, \vrms the \rms velocity in the convection zone, and \rib
the bulk Richardson number [see \cref{eq:rib}]. For the results presented in
the following, we have checked by visually inspecting the time evolution of the
density and boundary profiles that $\ve$ is dominated by mass entrainment and
the impact of thermal expansion is negligible. The analysis with respect to
\cref{eq:bri-scaling} is reported in various other studies which generally
find an agreement with the measured entrainment (e.g., \citealp{gilet2013a},
\citealp{mueller2016a}, \ci, \cii, \citealp{higl2021a}). In the following,
we extend these studies for the case of helium shell burning.

By fitting \cref{eq:bri-scaling} to simulations of mixing across boundaries at
different \rib, the value of $A$ and $n$ can be determined. In shell
simulations, this is possible either by measuring the entrainment at the bottom
and top boundary in a single simulation (different \rib because of different
\bvf profiles, e.g., \ci), by measuring entrainment in simulations with
different convective driving (different \rib because of varying \vrms), or both
(e.g., \cii).

To extract meaningful results, such simulations need to be run for multiple
convective turnover times.  Furthermore, as pointed out by
\citet{woodward2015a}, the resolution must be sufficiently high for obtaining
converged entrainment rates across the boundaries. To relate the different grid
sizes that have been used in the aforementioned studies, we compare the number
of vertical grid cells ($\ncz$) that are located within the convection zones.
Only grids that have been used to derive an entrainment rate are considered
here. This simple comparison neglects the impact of restricted domains and does
not consider the different stiffness of transitions from stable to convection
zones. However, it still gives an estimate of the scales that are resolved by
the grid compared to the global scale of the convection zone.
\citet{woodward2015a} find entrainment rates that are in reasonable agreement
for simulations with $\ncz=219$ (grid sizes of $768^3$) and more cells. Most of
the simulations presented by \citet{jones2017a} have $\ncz=170$, while they
show for one particular case that entrainment agrees with the results of a
simulation with $\ncz=341$ (grid sizes of $768^3$ and $1536^3$, respectively).
The highest resolution used by \ci to determine the entrainment rate has
$\ncz=256$ (for a grid of $512^3$). Our computational resources only allow to
run simulations long enough on grids with \lres cells which corresponds to
$\ncz=105$. This resolution might not be sufficient and we cannot test whether
the results presented in this section are converged. However, our analysis
still provides a first glimpse on what coefficients might be expected for
the He-shell burning. Moreover, we are able to compare the results from the
low-Mach \ausmpup solver to \ausmp and assess whether the bulk Richardson
scaling can be reproduced even at low resolution.

We determine the entrainment rate $\ve$ in \cref{eq:bri-scaling} from the mean
radial position over time of the top and bottom boundary, respectively. The
positions of the boundaries are extracted as described in \cref{sec:pstrace},
the values for \vrms consider the entire convection zone.

\begin{figure*}
  \centering
  \includegraphics[width=\textwidth]{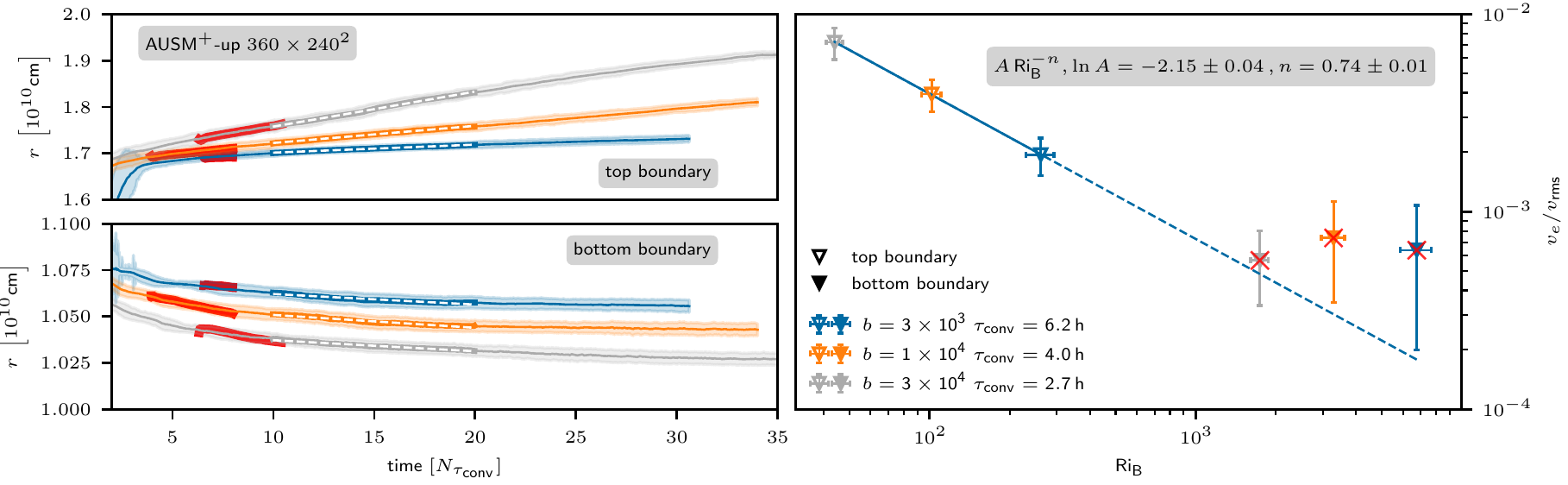}
  \caption{\textbf{Left plots:} Radial positions of the top and bottom
  boundaries as functions of multiples of the respective convective turnover
  times. The radial position is the average position over all angles, the
  shading marks the corresponding spatial standard deviation of the boundary
  position. The white dashed lines indicate the time frame considered to obtain
  the data for the right column. Thick red lines show the evolution of
  simulations with a grid size of \mrest cells for comparison. The scales of
  the $y$-axis for the top and bottom boundary have been adapted for better
  visibility. All simulations apply the \ausmpup scheme while the nuclear
  energy release is boosted by factors of \numlist{3e3;1e4;3e4}.
  \textbf{Right plot:} ratio of entrainment velocity \ve and \rms velocity
  \vrms as a function of the bulk Richardson number \rib for the simulations
  shown in the left column. The horizontal error bars indicate the standard
  deviations of the temporal means of \rib. Vertical error bars correspond to
  the standard deviations of the entrainment velocities for a sliding window of
  size $\Delta\Ntconv=1$. The time frame is large enough to filter out sound
  waves but it is somewhat arbitrary. However, the error bars give an idea of
  the spread in the entrainment velocity. Markers that are crossed are not
  considered for fitting \cref{eq:bri-scaling} to the data. The fit is shown as
  solid blue line. The dashed line at high \rib marks the regime of
  extrapolation.}
  \label{fig:line_bpos_scaling}
\end{figure*}
\begin{figure*}
  \centering
  \includegraphics[width=\textwidth]{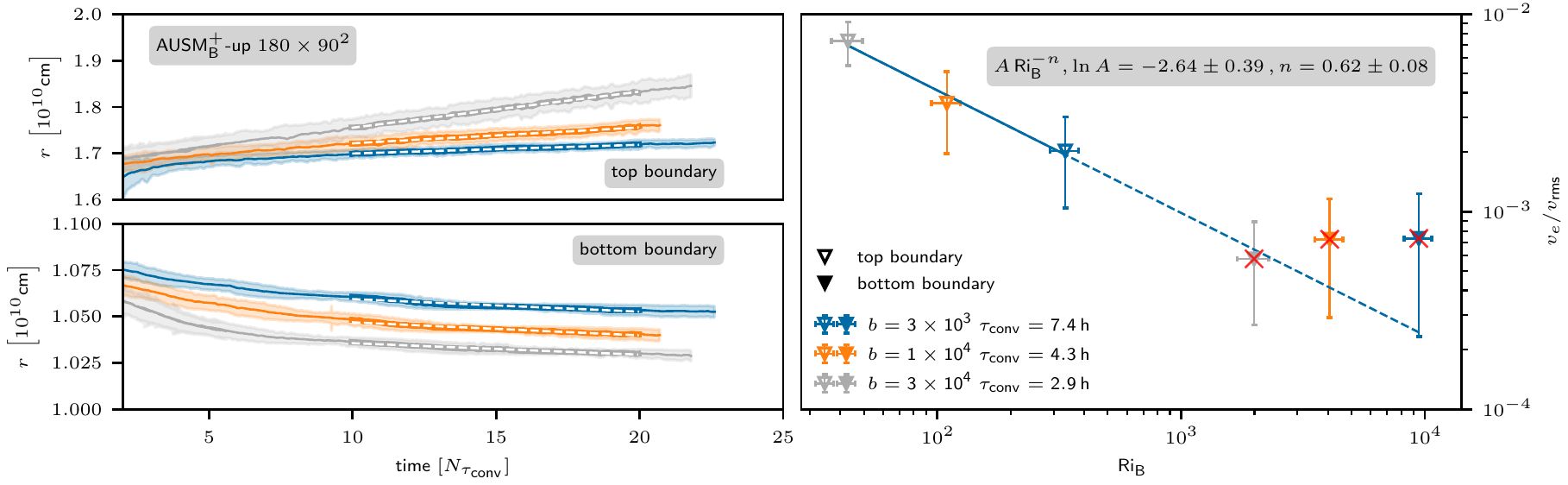}
  \caption{Same as \cref{fig:line_bpos_scaling} but for simulations using the \ausmp solver.}
  \label{fig:line_bpos_scaling_basic}
\end{figure*}
The plots on the left of \cref{fig:line_bpos_scaling} show the evolution of the
boundary positions for the boosting factors \numlist{3e3;1e4;3e4} when using
the \ausmpup solver. Qualitatively, the behavior is as expected: A higher
boosting factor leads to stronger convection, faster entrainment of the passive
scalar, and thus to a faster moving boundary. The entrainment velocity at the
bottom boundary is considerably smaller than at the top boundary.  According to
\cref{eq:bri-scaling} this is expected for a larger value of \rib which is
indeed confirmed in the right panel, where the data for the bottom boundary
reside at $\rib>\num{e3}$. A similar situation is observed for simulations with
the \ausmp solver, shown in \cref{fig:line_bpos_scaling_basic}.

\begin{table*}
  \centering
  \caption{Summary of the basic properties of the simulations
  shown in \cref{fig:line_bpos_scaling,fig:line_bpos_scaling_basic}.}   \label{tab:evel}
  \begin{tabular}{lcccccc}
    \toprule boosting factors        &
    \multicolumn{2}{c}{\num{3e4}} & \multicolumn{2}{c}{\num{1e4}} & \multicolumn{2}{c}{\num{3e3}} \\
    num.\ flux function      & \ausmpups & \ausmps & \ausmpups & \ausmps & \ausmpups & \ausmps  \bspace\tspace \\
    \hline
    $\tau_\text{conv}\, \left[\si{\hour}\right]$            &  2.69 &  2.86 &  3.96 &  4.32 &  6.21 & 7.40 \bspace\tspace  \\
    $\ve^B\, \left[\SI{e2}{\centi\meter\per\second}\right]$ & -6.14 & -6.05 & -4.86 & -4.52 & -2.54 & -2.55 \bspace\tspace \\
    $\ve^T\, \left[\SI{e3}{\centi\meter\per\second}\right]$ &  7.80 &  7.60 &  2.60 &  2.20 &  0.77 & 0.71 \bspace\tspace  \\
    $\ve^T/\vds$                                            &  0.45 &  0.54 &  0.39 &  0.52 &  0.34 & 0.41 \bspace\tspace  \\
    $\vpc^B\,\left[\text{cell}\,\tconv^{-1}\right]$         & -0.10 & -0.10 & -0.11 & -0.12 & -0.09 & -0.11 \bspace\tspace \\
    $\vpc^T\,\left[\text{cell}\,\tconv^{-1}\right]$         &  1.23 &  1.28 &  0.61 &  0.56 &  0.28 & 0.31 \bspace\tspace  \\
    $\vrms\, \left[\SI{e6}{\centi\meter\per\second}\right]$ &  1.08 &  1.04 &  0.66 &  0.62 &  0.40 & 0.35 \bspace\tspace  \\
    $\rib^B\, \left[\num{e3}\right]$                        &  1.74 &  1.99 &  3.29 &  4.07 &  6.76 & 9.44 \bspace\tspace  \\
    $\rib^T\, \left[\num{e2}\right]$                        &  0.44 &  0.43 &  1.02 &  1.09 &  2.62 & 3.35 \bspace\tspace  \\
    \bottomrule
  \end{tabular}
  \tablefoot{ All data is obtained by considering a time interval of
  $\Delta\Ntconv=10$.  Legend: $\tconv:$ mean convective turnover
  time. $\ve^B,\,\ve^T:$ entrainment velocities at the bottom and top
  boundary. $\ve^T/\vds$: Ratio of the entrainment velocity at the top
  boundary to the velocity estimated by a general entropy increase
  within the convection zone, see text and \cref{eq:vds}.
  $\vpc^B,\,\vpc^T:$ number of vertical grid cells crossed by the
  bottom (top) boundary over the period of one \tconv.  $\vrms:$ \rms
  velocity in the convection zone.  $\rib^B,\, \rib^T:$ bulk Richardson number
  at the bottom and top boundary.}
\end{table*}
In order to fit \cref{eq:bri-scaling} to the data, we consider the time frame
of $t\in\left[\tofn{10},\tofn{20}\right]$. The lower limit is given by the end
of the initial transient in the evolution of the passive scalar. The length of
the shortest simulations constitutes the upper limit, such that the same time frame
can be used for both sets of simulations. The extracted values listed in
\cref{tab:evel} reveal that the bottom boundaries move only by about
one cell during the entire considered time frame of $\Delta\Ntconv=\num{10.0}$.
This indicates that our grid is not fine enough to properly track this subtle
shift, which is also suggested from the thin boundary widths measured for the
bottom boundary, see \cref{sec:bwidth}. An additional complication arises by the
profile of energy generation (\cref{fig:line_initprofs}) which has its peak
beneath the convection zone.  Because we do not increase the efficiency of
radiative diffusion in accordance with the artificial boosting, internal energy
will accumulate below the convection zone. This leads to a local increase in the
\bvf and the boundary gets stiffer. Hence, the entrainment velocity decreases
when the bottom boundary approaches the peak of the energy generation. Because
of this artificial phenomenon and the unresolved boundary motion we exclude
the data points of the bottom boundaries from the analysis.

Using a least-square fit of \cref{eq:bri-scaling} to the extracted data, we find
\begin{align}
  \ln A_{\apup} &= \num{-2.15}\pm0.04,\, &n_{\apup} =\num{0.74}\pm 0.01, \nonumber\\
  \ln A_{\aup}  &= \num{-2.64}\pm0.39,\, &n_{\aup}  =\num{0.62}\pm 0.08,
  \label{eq:bri-res1}
\end{align}
where the errors correspond to the standard deviation of the fitting
parameters. We note that the errors are obtained without taking the individual
error bars shown in \cref{fig:line_bpos_scaling,fig:line_bpos_scaling_basic}
into account. The standard deviation in \rib and the spread in the entrainment
velocity are likely correlated between some of the data points and subject to
systematic shifts. Therefore, we think a treatment in terms of proper
error propagation could be misleading. The error bars are shown nonetheless in
the figures to give an idea of the general variability of the data points. The
small uncertainties given in \cref{eq:bri-res1} thus indicate only that our
data is well represented by \cref{eq:bri-scaling} but should not be taken as a
measure of the overall accuracy of our analysis.

The results with the \ausmpup and \ausmp schemes are similar, but
the dependence of the entrainment on the bulk Richardson number is somewhat
steeper for \ausmpup. As a rough test for convergence, the set of simulations
with the \ausmpup solver has been restarted after the initial transient on
grids with a resolution of \mrest. The flow state is interpolated to the finer
grids using constant interpolation. The corresponding tracks of the radial
boundary positions are shown as thick red lines in
\cref{fig:line_bpos_scaling}. For the top boundary, the entrainment rate is
similar to the low resolution runs. At the bottom boundaries, entrainment
appears to be slightly faster. Generally, the better resolved simulations
follow a similar trend as the low resolution runs. However, more data is needed
for a stronger statement on convergence and to extract meaningful estimates for
$A$ and $n$ also from the better resolved simulations.
\begin{figure}
  \centering
  \includegraphics[width=\columnwidth]{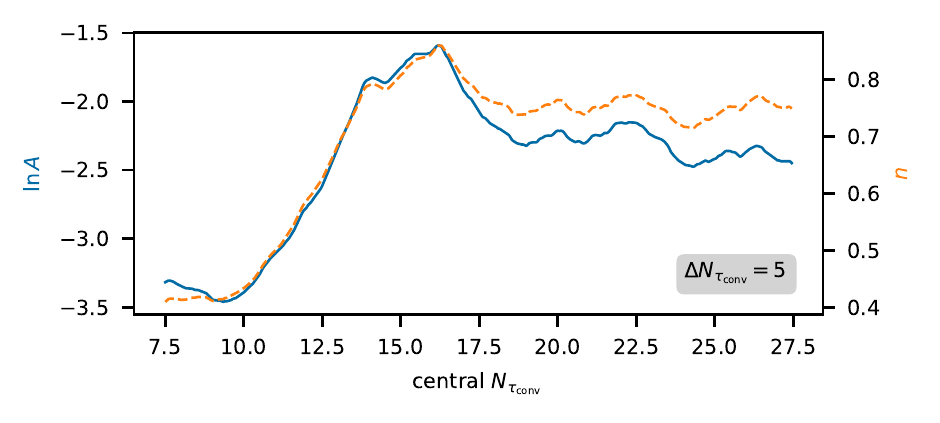}
  \caption{Time evolution of the fit parameters $A$ (blue solid line) and $n$
  (orange dashed line) for a fixed time frame of $\Delta\Ntconv=5$ and moving
  central time.}
  \label{fig:line_timeevo_move}
\end{figure}
Another parameter that impacts the results is the considered time interval.
Using the full data of the \ausmpup runs shown in \cref{fig:line_bpos_scaling},
we extract the parameters $A$ and $n$ for a fixed length of $\Delta\Ntconv=5$
but for a changing central time (\cref{fig:line_timeevo_move}). We find that
the value of the exponent $n$ increases from $n\approx\num{0.4}$ and settles to
a value of $n\approx\num{0.75}$ for central times later than \tofn{20}. The
value of $\ln A$ changes in a very similar way from $\ln A\approx\num{-3.5}$ to
$\ln A\approx\num{-2.5}$. \cref{fig:line_timeevo_move} reveals that  the values
settle after central $\Ntconv \approx 20$.  Therefore, it seems more
appropriate to consider the time interval of
$t\in\left[\tofn{17.5},\tofn{25}\right]$ to determine best-fitting values of
$A$ and $n$. The upper limit is given by the time the top boundary
reaches the top of the computational domain and boundary conditions will start
to affect the results. We obtain
\begin{align}
  \ln A_{\apup} &= \num{-2.24}\pm0.45,\, &n_{\apup} =\num{0.76}\pm 0.10,
  \label{eq:bri-res2}
\end{align}
see also \cref{fig:line_bpos_scaling_compare}. These values are similar to the
previous result. However, considering the peak in the evolution in
\cref{fig:line_timeevo_move} between $\Ntconv=10$ and $\Ntconv=20$, this might
be a coincidence. There is not sufficient data for the simulations with the
\ausmp solver to repeat this calculation but we expect it to show a similar
trend.

The results given in \cref{eq:bri-res1,eq:bri-res2} are within the regime
$\num{0.5}\le n \le 1$, which is compatible with values reported in laboratory
and numerical experiments (see, e.g., \citealp{meakin2007b} and \ci for a
discussion and corresponding references). Our fitting parameters  are similar
to the findings of \cii for the carbon shell. They obtain
parameters\footnote{We note that the value and error of $A$ given in their
\figr{14} mix linear (for A value) and logarithmic (for the uncertainty)
scales. The values presented here are recalculated from the same data set in
terms of the natural logarithm.} of $\ln A_\text{C19} = \num{-2.98}\pm 0.13$
and $n_\text{C19} =\num{0.74}\pm\num{0.04}$. In contrast, \citet{meakin2007b}
report $\ln A_\text{M07} = \num{0.062}\pm\num{0.87}$ and $n_\text{M07}
=\num{1.05}\pm\num{0.21}$ and also the results of \citet{jones2017a} and
\citet{andrassy2020a} agree with an exponent of $n\approx1$, as pointed out by
\citet{mueller2020a}. Further simulations are needed to scrutinize the values
of $A$ and $n$, also keeping in mind that different values may exist
for different stellar convection zones.

Combining the results of \cref{eq:bri-res2} with the \mlt prediction of
$\ma_{\mlt} \approx \num{e-4}$ and $\rib = \num{7e4}$ for the 1D \mesa model,
we find a mass entrainment rate of
\begin{align}
  \dot{m}_e &= 4\pi r^2 \rho\, \ma_{\mlt}\,\csound\, A \rib^{-n} \nonumber\\
            &= \SI{9.6e-11}{\msol\per\second},
  \label{eq:mass_entr}
\end{align}
for the top boundary. The value for the bottom boundary is about a factor ten
smaller. This confirms the finding of previous 3D hydro simulations (e.g., \ci)
that lower boundaries of convective shells are stiffer and thus have less
entrainment than the top boundaries. The associated growth of the convective
region at the upper and lower boundaries using these rates until the end of the
evolution is indicated by green lines in \cref{fig:kipp_initial}. This
illustrates that, while the much stiffer lower boundary only slightly changes,
the upper boundary considerably moves outward. At the rate of the mass
entrainment of \cref{eq:mass_entr}, the total mass of the initial convection
zone of \SI{1.1}{\msol} is doubled within \SI{350}{\year}.  However, a
substantial growth of the convection zone leads to different bulk Richardson
numbers at the boundaries and thus change the mass entrainment rate.  Moreover,
as seen in \cref{fig:kipp_initial}, the convection zone is growing also in the
1D evolution calculation. The mass entrainment rate given in
\cref{eq:mass_entr} is therefore only representative for a short fraction of
the shell's lifetime at the evolutionary time the simulation was calculated. It
is also a warning that one cannot simply use numerical values like entrainment
rates extracted from single 3D simulations and apply them at different phases of
stellar evolution. Instead, it is best to use theoretical prescriptions like
the entrainment law. Recently, 1D stellar evolution models using the
entrainment law on the main-sequence have been computed by \citet{scott2021a}
and better reproduce the mass dependence of the main-sequence width. New 1D
models during other phases of stellar evolution will be needed to assess the
ability of the entrainment law to represent convective boundary mixing in 1D
models throughout stellar evolution.

The expansion of the convection zone seen in the 1D evolution is part of a
global restructuring of the star after core helium burning. Turbulent
entrainment does not contribute as it is not included in our 1D calculation.
However, the growth may be attributed to a process that is likely also
present in our hydrodynamic simulations: In a simplified picture, the heating
through nuclear burning successively increases the entropy within the
convection zone. This leads to a small region at the top boundary where it
exceeds the entropy at the immediate beginning of the radiation zone. This
region is unstable and will merge with the convection zone. We estimate the
speed $\vds$ at which this process would move the outer boundary by
\begin{align}
  \vds = {\left({\frac{\Delta s_{\text{CZ}}}{\Delta t}}\right)}\, /\,          {\left({\frac{\Delta s_{\text{RZ}}}{\Delta r}}\right)},
  \label{eq:vds}
\end{align}
where $\Delta s_{\text{CZ}} / \Delta t$ is the ratio of the mean temporal increase
in entropy inside the convection zone and $s_{\text{RZ}}/\Delta r$ corresponds
to the mean entropy gradient at $t=0$ for a region above the top boundary.
\begin{figure}
  \centering
  \includegraphics[width=\columnwidth]{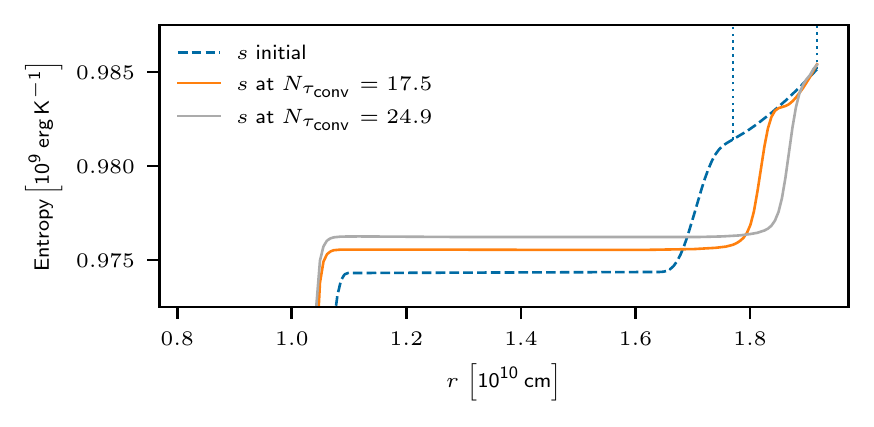}
  \caption{Initial entropy profile and profiles after $17.5$ and $25$
  convective turnover times for the 3D simulation with a grid of \lres cells,
  the \ausmpup solver, and an energy boosting of $b=\num{3e4}$. Vertical dotted
  lines mark the region that is considered to calculate the mean entropy gradient
  in the radiation zone $\Delta s_{\text{RZ}}/\Delta r$ from the initial
  entropy profile.}
  \label{fig:line_grads_estimate}
\end{figure}
For the simulation with the highest boosting that was used to obtain the
results in \cref{eq:bri-res2} we find a ratio of
\begin{align}
   \frac{\vds}{\ve^T} \approx \SI{60}{\percent},
\end{align}
where $\ve^T$ is the entrainment velocity at the top boundary as measured from
the advected passive scalar. The considered profiles to calculate \vds are
plotted in \cref{fig:line_grads_estimate}, the spatial entropy gradient is
calculated considering the initial model. The ratios of our other simulations
range between $\SI{30}{\percent}$ to $\SI{50}{\percent}$ and are listed in
\cref{tab:evel}. This is similar to the value of \SI{49}{\percent} found by
\citet{andrassy2020a} for carbon-shell burning while \citet{meakin2007b} find a
maximum ratio of \SI{17}{\percent} for oxygen shell burning\footnote{It is not
clear to us whether \citet{meakin2007b} calculated the spatial entropy
gradient in the radiation zone or at the transition from convection to
radiation zone. In the latter case, the gradient is much steeper, the
estimated velocity will be smaller, and we would obtain a ratio similar
to that of \citet{meakin2007b}. However, we think that only the
gradient above the boundary transition is relevant.}. These values
suggest that a considerable fraction of the entrainment speed could be
contributed through increasing entropy.  Consequently, this process
needs to be disentangled from the growth through turbulent entrainment
before the entrainment according to \cref{eq:bri-scaling} is used in
stellar evolution codes or compared between simulations of different
convection zones.

\subsection{Characterizing the mixing}
\label{sec:chm}

It is not trivial to determine the details of the -- possibly  small-scale --
events that lead to turbulent mixing. In their PPMstar simulations,
\citet{woodward2015a}, \citet{jones2017a}, and \citet{andrassy2020a} find that
trains of small Kelvin-Helmholtz rolls emerge at the boundary. However, they
do not appear in regions of largest shear but rather at the point where two
convective cells turn and move back into the convection zone.

From our 2D visualizations we are not able to easily find large scale, coherent
modes. In order to identify possible correlations between the strength of
shear and the amount of mixing in our simulations, we use a simple analysis of
the velocity field at the top boundary: For a narrow region below the top
boundary, we measure along radial rays the volume-weighted deviation of the
passive scalar from its mean in the convection zone
\begin{align}
  \widetilde{\text{PS}}(\vartheta,\varphi) =
    \frac{\sum_{r\in\left[r_{\text{PS}},r_{\text{P}}\right]}
    V(r,\vartheta,\varphi)\left[\text{PS}(r,\vartheta,\varphi)-\overline{\text{PS}}(\vartheta,\varphi)\right]}
    {\sum_{r\in\left[r_{\text{PS}},r_{\text{P}}\right]} V(r,\vartheta,\varphi)},
  \label{eq:psfluc}
  \end{align}
where $\text{PS}$ denotes the value of the passive scalar,
$\overline{\text{PS}}(\varphi,\vartheta)$ is the average over the middle third
of the convection zone, and $V$ the volume of the grid cell. The radii
$r_{\text{PS}},\,r_{\text{P}}$ define the considered radial domain,
where $r_{\text{P}}$ corresponds to the beginning of the transition to the stable
zone at the top of the convection zone, as defined in \cref{sec:bwidth} and
$r_{\text{PS}}=0.95\,r_{\text{P}}$. The value of the passive scalar is larger
above the top boundary compared to its mean (see \cref{fig:line_cb_tracing}).
Hence, a positive deviation from the mean corresponds to an entrainment event.
The considered domain does not include mixing directly at the boundary because
there, deviations are usually large but do not necessarily descend into the
convection zone.

In addition, we estimate the strength of shear by
\begin{align}
  S_h(\vartheta, \varphi) = \int_{r_{\text{PS}}}^{r_{S}} \sqrt{\left(\partial_r v_\varphi\right)^2 + \left(\partial_r v_\vartheta\right)^2}\,\dd r,
  \label{eq:shear}
\end{align}
where $v_\vartheta,v_\varphi$ denotes the $\vartheta,\varphi$-velocity
components. Because the shear at the boundary matters here, we extend the
considered zone to $r_{S}$ which coincides with the end of the transition to
the radiation zone as defined in \cref{sec:bwidth}.
\begin{figure}
  \centering
  \includegraphics[width=\columnwidth]{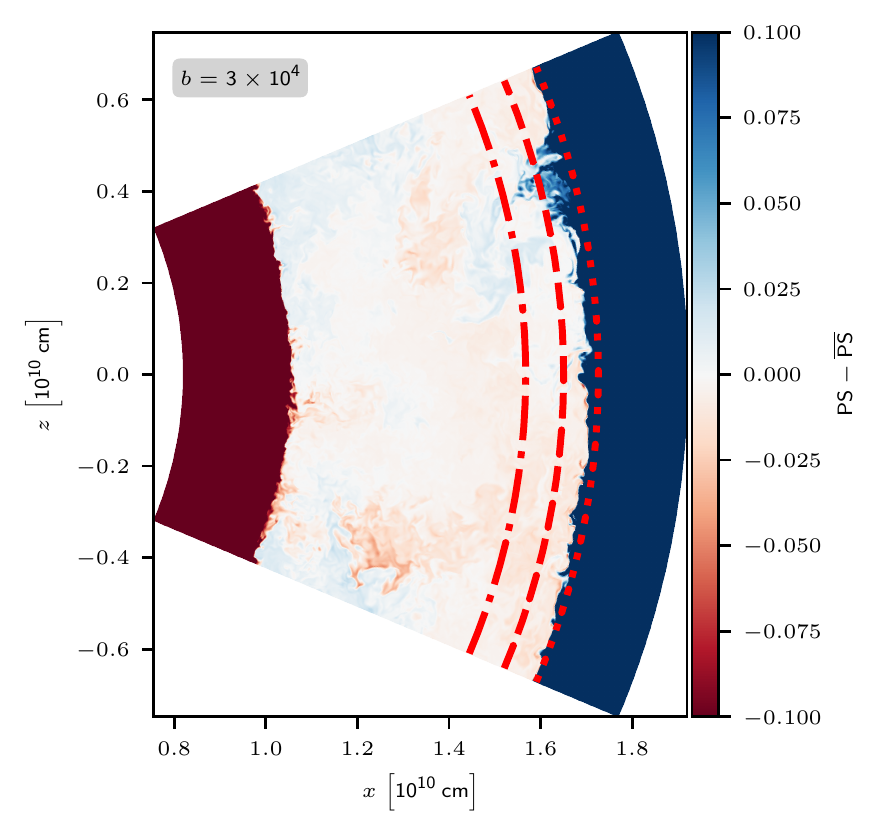}
  \caption{Fluctuations of the passive scalar PS from its mean value
  $\overline{\text{PS}}$ for a snapshot of the 3D simulation at a resolution
  of \vhres cells and a boosting factor of \num{3e4}.  The mean
  $\overline{\text{PS}}$ is taken over the inner third of the convection zone.
  The dashed-dotted line corresponds to $r_{\text{PS}}$, the dashed line to
  $r_{\text{P}}$ and the dotted line to $r_{S}$. Their meanings are explained
  in the main text.}
  \label{fig:pcolor_shear_mix_ill}
\end{figure}
The different regions are indicated in \cref{fig:pcolor_shear_mix_ill}. With
the described procedure we obtain data pairs that correlate shear
strength to mixing strength.
Our simple approach does not consider that the mixing events will also
depend on the history of the velocity field and its gradient along the
individual downflows. However, it still gives some measure of the correlation
between shear and mixing: The characteristic time scale for global changes of
the flow field is given by the turn over time. The animated versions\footnote{
\href{https://zenodo.org/record/4776452}{https://zenodo.org/record/4776452}} of
\cref{fig:pcolor_shear_mix_ill} illustrate that the
mixing events detected between the dashed-dotted line and the dashed line
happen on time scales which are much shorter. If the mixing were caused by
Kelvin-Helmholz instabilities overturning the whole boundary layer one may
assume that they grow fastest in regions of strongest shear. We then would
expect the rapidly-growing Kelvin-Helmholz rolls to become detectable in the
layer where we measure $\widetilde{\text{PS}}$ after a fraction of the global
turnover time scale. However, the shear layers caused by the overturning of the
large scale flows at the convective boundary should be present for as long as
the global turnover time scale. The extracted data pairs can therefore be used
to investigate a possible correlation between shear strength and mixing.

\begin{figure}
  \centering
  \includegraphics[width=\columnwidth]{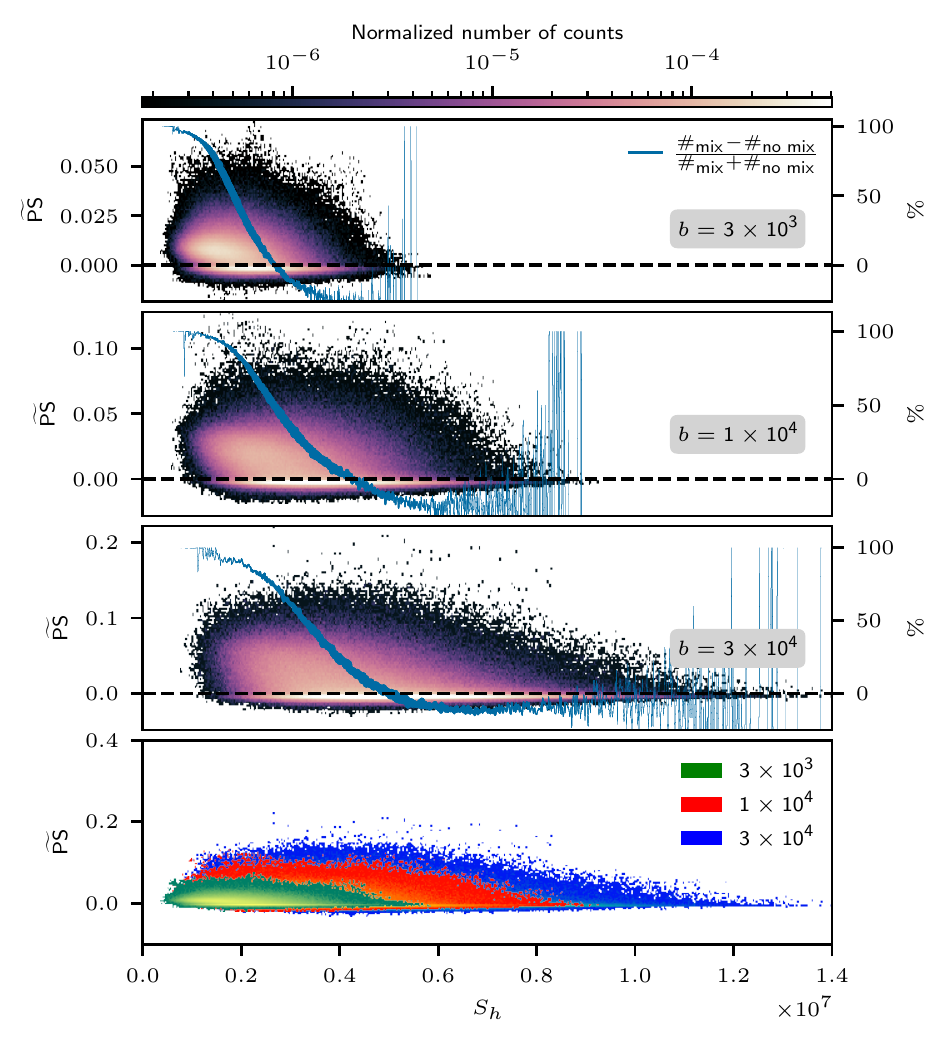}
  \caption{Histogram of fluctuations of the passive scalar $\tilde{\text{PS}}$
  [\cref{eq:psfluc}] as a function of shear strength for simulations with the
  \ausmpup solver and a resolution of \lres. The boosting strength increases
  from top to bottom. The lowest panel combines all three histograms
  to ease the direct comparison. The values of fluctuations and shear
  strengths are extracted in a narrow band below the top boundary of the
  convection zone, as illustrated in \cref{fig:pcolor_shear_mix_ill}. The
  histograms are normalized by the total number of counts, that is the number
  of horizontal grid cells multiplied by the number of considered grid files.
  The blue lines indicate the relative excess of mixing events (positive
  passive scalar fluctuation) or ``no mixing'' events (negative passive scalar
  fluctuation) relative to the total number of events for the respective shear
  strength. The width of the lines is scaled linearly with the relative
  contribution of the counts at the respective shear strength to the global
  number of counts. A thick line therefore indicates a significant contribution
  while the thin lines at very small and large shear strengths indicate a
  negligible contribution to the total amount of events. The considered time
  frame is $t\in\left[\tofn{17.5},\,\tofn{25}\right]$.}
  \label{fig:pcolor_shear_mix_low}
\end{figure}

The result for the \ausmpup solver at a resolution of \lres is shown in
\cref{fig:pcolor_shear_mix_low} for increasing energy boostings. In all
simulations, the counts of positive passive scalar fluctuations cluster at the
lower end of the measured shear strength range. At slightly negative
deviations, a narrow band with a high number of counts forms which stretches
over a larger range of shear strengths. As indicated by the blue lines,
at smaller shear strengths mixing events dominate over ``no mixing'' events.
The noisy profile at the strongest shear can be attributed to the small number
of total counts (thin line) and corresponding poor statistics.

The evidence of mixing at the lower end of the range is in
line with the fact that the horizontal velocity naturally has to decrease where
strong downflows form because the velocity is redirected inward there. The
narrow band corresponds to rays with no mixing events such that the contained
passive scalar is slightly below but very close to the average within the
convection zone. If the energy boosting is increased, convection gets more
vigorous and hence the narrow band extends toward larger shear strengths.
Mixing follows this trend, but still dominantly appears at lower shear
strengths. By visually inspecting the flow morphology of their 3D
simulations, \citet{woodward2015a} find that mixing predominately occurs in
regions where two large convective cells meet and overturn. The premixed
material that accumulates in the wedge between two cells somewhat beneath the
boundary has a sufficiently small buoyancy force with respect to the bulk of
the convection zone such that the downflows are strong enough to bring the
material inward. Because of the decreasing horizontal velocity of the turning
cells, this premixed region will necessarily have weaker shear [as measured by
\cref{eq:shear}] compared to the region where the fluid moves almost horizontally.
The results of our analysis seem to support this picture.

\begin{figure}
  \centering
  \includegraphics[width=\columnwidth]{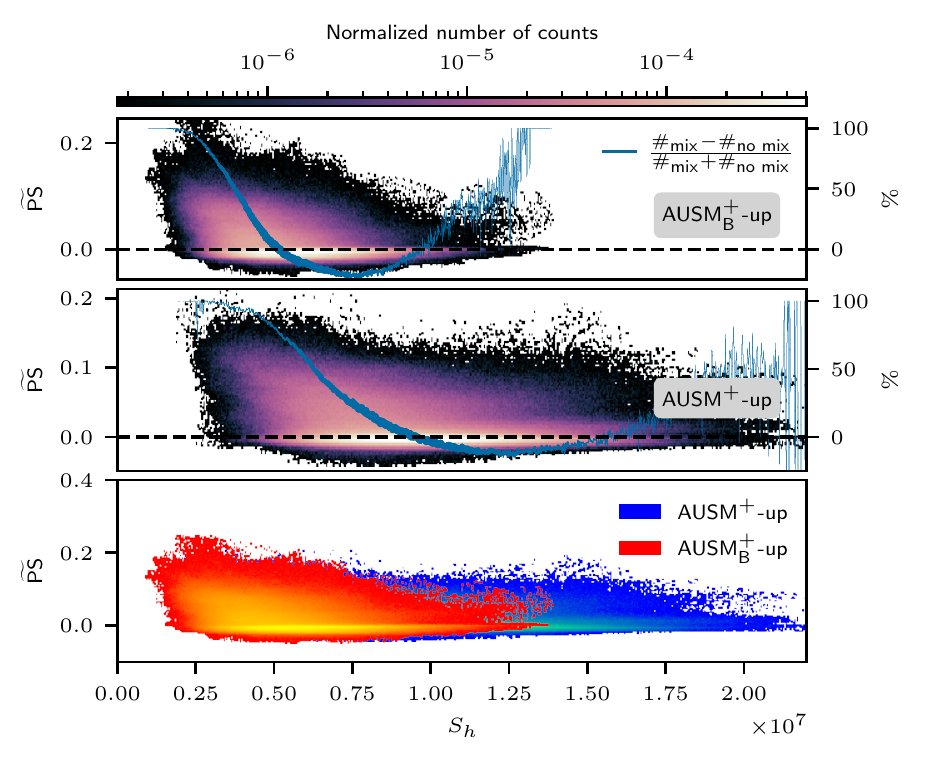}
  \caption{Same as \cref{fig:pcolor_shear_mix_low}, but here
  for the \ausmp (top panel) and \ausmpup (bottom lower) solver at a resolution
  of \vhres and an energy boosting of \num{3e4}. The considered time frame
  spans over $\Delta\Ntconv=0.5$, including the respective latest snapshot of
  each run.}
  \label{fig:pcolor_shear_mix_hres}
\end{figure}
In \cref{fig:pcolor_shear_mix_hres} we compare in a similar histogram the
results of the \ausmpup and \ausmp solver at the runs with highest resolution.
The direct comparison shows that shear values spread over a larger range for
the \ausmpup solver, which can be attributed to its better-resolved turbulent
flow (see \cref{fig:pcolor_flopat,fig:spechres}). The \ausmp solver shows
slightly stronger mixing events. The apparent return toward positive
deviations at large shear is insignificant due to the small number of total
counts at larger shear, as indicated by the thin line.

\subsection{Boundary width}
\label{sec:bwidth}
Another characteristic of convection is the shape of the boundary layers
between the convection zone and the convectively stable zones. While the
boundary width has to be parametrized in 1D stellar evolution codes, it
develops self-consistently in hydrodynamic simulations. As can be seen in
\cref{fig:line_cb_tracing}, the transition that forms during the simulation is
not sharp but rather changes gradually across a certain vertical width. This is
due to at least two processes. The first, and most important, is partial mixing
across the boundary layer by turbulent entrainment. The second, which is less
important in the simulations presented in this study, is the deformation of
the boundary layer producing an undulated surface rather than a perfectly
spherical surface. Neither of these processes exist in most 1D models, which
generally assume a sharp step-like boundary. The exception is 1D models using
partial mixing beyond the convective boundary such as the prescriptions of
exponentially decaying mixing by \citet{freytag1996a} and \citet{herwig1997a}.

Comparisons to asteroseismology (e.g.,
\citealp{moravveji2016a,arnett2017a,michielsen2019a}) also support smoother
over step-like boundaries. More work is needed to better understand the shape
boundaries since they can have a decisive impact on the evolution and
nucleosynthesis \citep[e.g.,][]{battino2016a}.

In this section we compare the transition layer widths for simulations with
different resolutions, flux solvers and boostings. Our approach to extract the
widths is similar to the procedure described by \ci but instead of abundance
profiles we use the passive scalar as tracer. We define the inner radii of the
transitions at the bottom (top) of the convection zone as the radii at which
the horizontal mean of the passive scalar is larger than its initial value at
this radius plus (minus) \num{0.05}. The outer radius of the transitions at the
bottom (top) is taken to be the radius at which the horizontal mean of the
passive scalar is below (above) the spatial mean over the inner third by
\num{0.05}.  To determine the corresponding radii, the profile of the passive
scalar is interpolated. The resulting widths are shown exemplarily in
\cref{fig:line_cb_tracing}, marked by vertical dashed lines. The absolute
value of the width depends on the particular choice of the thresholds for the
deviations from the initial profile. However, it still gives a measure for the
relative dependence on resolution, boosting strength and numerical flux
solver. We have verified that the trends found for the
boundary widths do not depend on the specific choice of the threshold value.

\begin{table*}
  \centering
  \caption{Boundary widths of the bottom and top boundaries for different
  energy boosting factors at a resolution of \lres cells.} 
  \centering
  \begin{tabular}{lcccc}
  \toprule
               & \multicolumn{2}{c}{\ausmps}               &  \multicolumn{2}{c}{\ausmpups}                      \\
   boost               & $\delta_\text{r,\,bot}$                           &   $\delta_\text{r,\,top}$ &   $\delta_\text{r,\,bot}$ &   $\delta_\text{r,\,top}$ \\
                       & \multicolumn{2}{c}{{\footnotesize $\left[\SI{e8}{\centi\meter}\right]$}} & \multicolumn{2}{c}{{\footnotesize $\left[\SI{e8}{\centi\meter}\right]$}}\bspace\tspace   \\
  \hline
  $3.0 \times 10^{3}$ & $1.57 \pm 0.07$           & $5.27 \pm 0.22$           & $2.11 \pm 0.11$           & $3.70 \pm 0.05$ \tspace   \\
  $1.0 \times 10^{4}$ & $1.72 \pm 0.07$           & $6.51 \pm 0.37$           & $2.25 \pm 0.08$           & $4.51 \pm 0.12$ \bspace   \\
  $3.0 \times 10^{4}$ & $1.93 \pm 0.06$           & $8.88 \pm 0.37$           & $2.30 \pm 0.09$           & $5.71 \pm 0.33$ \bspace   \\

  \midrule
  \rule{0pt}{2.0ex}
                      & $\delta_\text{r,\,bot}$   &   $\delta_\text{r,\,top}$ &   $\delta_\text{r,\,bot}$ &   $\delta_\text{r,\,top}$ \\
                      & \multicolumn{2}{c}{{\tiny $\left[\SI{e-2}{H_p}\right]$}} & \multicolumn{2}{c}{{\tiny $\left[\SI{e-2}{H_p}\right]$}}\tspace\bspace \\
  \hline
  $3.0 \times 10^{3}$ & $5.58 \pm 0.25$           & $15.41 \pm 0.63$          & $7.52 \pm 0.40$           & $10.81 \pm 0.15$ \tspace  \\
  $1.0 \times 10^{4}$ & $6.17 \pm 0.24$           & $18.94 \pm 1.07$          & $8.11 \pm 0.29$           & $13.13 \pm 0.36$ \bspace  \\
  $3.0 \times 10^{4}$ & $7.01 \pm 0.23$           & $25.63 \pm 1.08$          & $8.40 \pm 0.31$           & $16.48 \pm 0.96$ \bspace  \\
  \bottomrule
  \end{tabular}
  \label{tab:widthboost}
  \tablefoot{The upper set shows the widths in units of \si{\centi\meter} while
   the lower set shows the widths in terms of the mean pressure scale
   height over the boundary width. The values are averages taken over a
   time interval of $\Delta\Ntconv=10$, starting at \tofn{10}. Errors
   correspond to the standard deviation of the temporal averages.}
\end{table*}
In \cref{tab:widthboost} the resulting widths are listed for simulations
applying the \ausmp and \ausmpup solver at a fixed resolution of \lres for
varying boosting strength. We find that generally the top boundary width is
larger than the bottom boundary. This is in line with the much higher bulk
Richardson number at the bottom boundary (\cref{tab:evel}). The top boundary
broadens with increasing energy input because stronger driving leads to
stronger convection and eventually to enhanced mixing that reaches further into
the stable zone. In addition, the interface gets more deformed. This is in
accordance with the results reported by \ci. Also the transition of the bottom
boundary widens with increasing driving and is generally narrower for runs with
the \ausmp solver. However, because of the stiffness of the bottom boundary,
changes are only subtle. With a radial grid spacing of about
\SI{0.6e8}{\centi\meter}, the bottom boundaries are resolved by a few cells
only. The relative changes are even on the subgrid level and can only be
followed by interpolation.  Hence, the robustness of these results is difficult
to assess.

\begin{figure}
  \centering
  \includegraphics[width=\columnwidth]{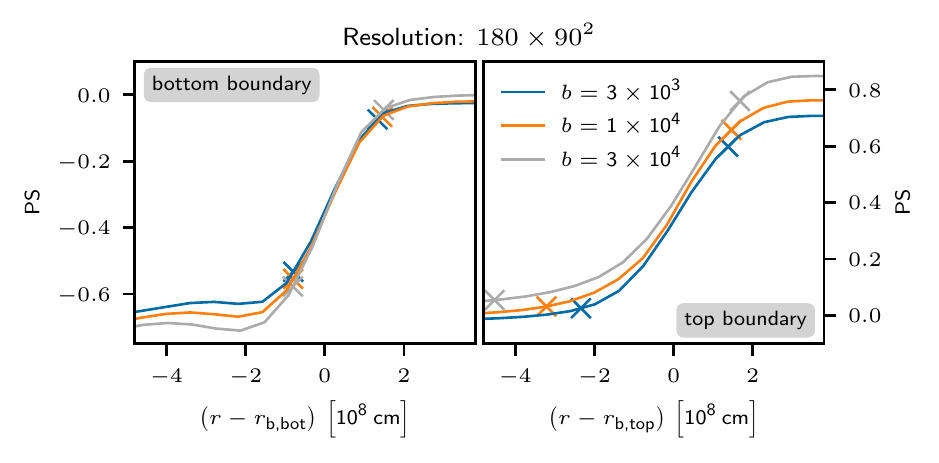}
  \caption{Profiles of the advected passive scalar for different energy input
  boostings. All simulations use the \ausmpup solver. The profiles are taken at
  \tofn{15}, crosses denote the beginning and end of the boundary transition
  zone, as defined in the text. Following the approach of \ci, the profiles are
  shifted by the radial position of the bottom ($r_\text{b,bot}$) and top
  ($r_\text{b,top}$) boundary, respectively.  The different amplitudes of the
  passive scalar below the bottom and above the top boundary are due to the
  fact that the initial profile is linear, see \cref{fig:line_cb_tracing}.
  Larger boosting leads to faster entrainment and the top boundary will have
  already moved toward larger radii, that is larger values of the passive
  scalar, for the snapshot shown in \cref{fig:line_bwidth_boost}.}
  \label{fig:line_bwidth_boost}
\end{figure}
\cref{fig:line_bwidth_boost} illustrates the boundary widths for one particular
point in time. This is similar to \figr{12} of \ci for the carbon-burning
shell simulations: For the top boundary, the broadening with increasing
energy input is clearly visible but it is less obvious at the bottom boundary.

\begin{figure}
  \centering
  \includegraphics[width=\columnwidth]{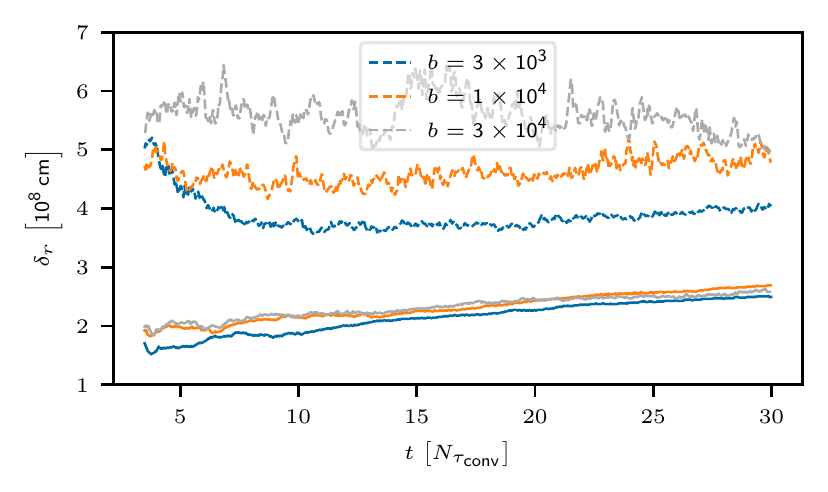}
  \caption{Widths of the upper (dashed lines) and lower (solid lines) boundary
  for different strengths of the energy generation boosting as a function
  of convective turnover times \tconv. All simulations use the \ausmpup flux.}
  \label{fig:bwidth_time}
\end{figure}
The time evolution of the boundary widths is shown in \cref{fig:bwidth_time}.
The upper transition exhibits some variability over time with an amplitude that
increases with the driving strength. Almost no fluctuations are visible for the
bottom boundary. A slight trend toward shallower transitions is noticeable.
These results confirm the general dependence of the boundary width on the
stiffness and driving strength.

\begin{table*}
  \centering
  \caption{Boundary widths of the bottom and top boundaries for different
  resolutions and a boosting factor of \num{3e4}.} 
    \begin{tabular}{lcccc}
    \toprule
                         & \multicolumn{2}{c}{\ausmp}                                               & \multicolumn{2}{c}{\ausmpup}  \\
     resolution          & $\delta_\text{r,\,bot}$    &   $\delta_\text{r,\,top}$                   &  $\delta_\text{r,\,bot}$ &   $\delta_\text{r,\,top}$ \\
                         & \multicolumn{2}{c}{{\footnotesize $\left[\SI{e8}{\centi\meter}\right]$}} & \multicolumn{2}{c}{{\footnotesize $\left[\SI{e8}{\centi\meter}\right]$}}\bspace\tspace\\
    \hline
	$180\times90^2$  & $2.10 \pm 0.03$           & $7.22 \pm 0.12$           & $1.94 \pm 0.04$           & $5.65 \pm 0.12$ \tspace   \\
	$360\times180^2$ & $1.93 \pm 0.08$           & $5.35 \pm 0.19$           & $1.38 \pm 0.03$           & $4.41 \pm 0.09$ \bspace   \\
	$540\times360^2$ & $1.73 \pm 0.04$           & $4.88 \pm 0.16$           & $1.26 \pm 0.04$           & $4.76 \pm 0.20$ \bspace   \\
	$810\times540^2$ & $1.49 \pm 0.04$           & $4.60 \pm 0.13$           & $1.07 \pm 0.03$           & $4.61 \pm 0.10$ \bspace   \\
    \midrule
    \rule{0pt}{2.0ex}
                         &   $\delta_\text{r,\,bot}$ &   $\delta_\text{r,\,top}$ &   $\delta_\text{r,\,bot}$ &   $\delta_\text{r,\,top}$      \\
                         &   \multicolumn{2}{c}{{\footnotesize $\left[\SI{e-2}{H_p}\right]$}} & \multicolumn{2}{c}{{\footnotesize $\left[\SI{e-2}{H_p}\right]$}}\tspace\bspace\\
    \hline
 	$180\times90^2$  & $7.52 \pm 0.12$           & $21.10 \pm 0.34$          & $6.96 \pm 0.16$           & $16.53 \pm 0.35$ \tspace  \\
 	$360\times180^2$ & $6.90 \pm 0.30$           & $15.64 \pm 0.55$          & $4.94 \pm 0.09$           & $12.87 \pm 0.27$ \bspace  \\
 	$540\times360^2$ & $6.17 \pm 0.14$           & $14.26 \pm 0.47$          & $4.48 \pm 0.16$           & $13.96 \pm 0.57$ \bspace  \\
 	$810\times540^2$ & $5.30 \pm 0.15$           & $13.44 \pm 0.39$          & $3.83 \pm 0.11$           & $13.47 \pm 0.29$ \bspace  \\
    \bottomrule
  \end{tabular}
  \label{tab:widthres}
  \tablefoot{Quantities have the same meaning as in \cref{tab:widthboost}.
  The values are averages over a time range of $\Delta\Ntconv =
  \num{0.5}$, the central time is \tofn{4.2}. Due to insufficient data,
  the central time is set to \tofn{2.9} for the run with the \ausmpup
  solver at a resolution of \hres cells.}
\end{table*}
\begin{figure}
  \centering
  \includegraphics[width=\columnwidth]{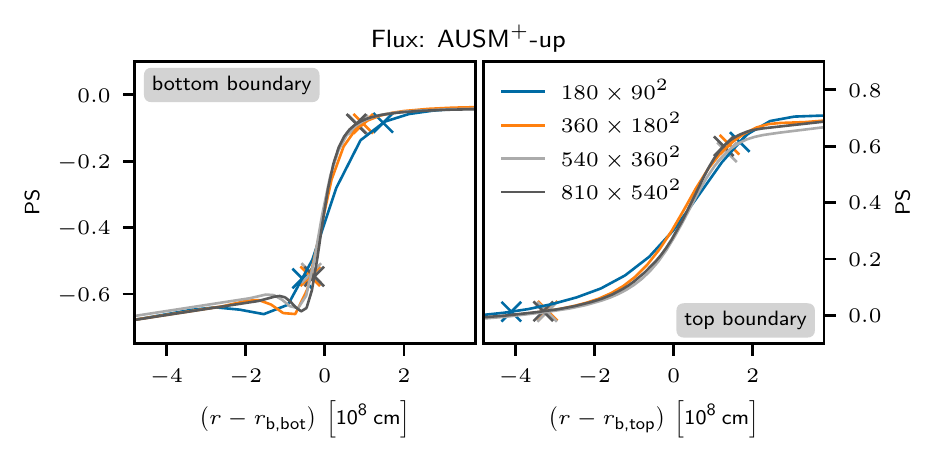}
  \caption{Similar to \cref{fig:line_bwidth_boost} but for a fixed energy
  boosting factor of \num{3e4} and varying resolution. All simulations use the
  \ausmpup solver. The profiles are taken at $\Ntconv=4.5$, except for the run
  with a grid of \hres cells. Here, the profile is taken at $\Ntconv=3.1$, the
  latest available snapshot.}
  \label{fig:line_bwidth_res}
\end{figure}
To assess the impact of resolution, we compare the widths at a boosting factor
of \num{3e4} for simulations on successively finer grids in
\cref{tab:widthres}. Because the finer resolved simulations cover less
convective turn over times, the averages are taken at earlier times compared to
the data listed in \cref{tab:widthboost}.

For both flux functions we find that the widths of the upper boundary
noticeably decrease when the grid is refined from a resolution of
$\num{180}\times\num{90}^2$ to $\num{360}\times\num{180}^2$. For even finer
grids, the width changes only slightly, which is confirmed in the boundary
profiles shown in \cref{fig:line_bwidth_res}. While the results seem to be
converged for the respective flux functions, there is still a discrepancy
between the solvers at the bottom boundary which persists even for the highest
resolution.
\begin{figure}
  \centering
  \includegraphics[width=\columnwidth]{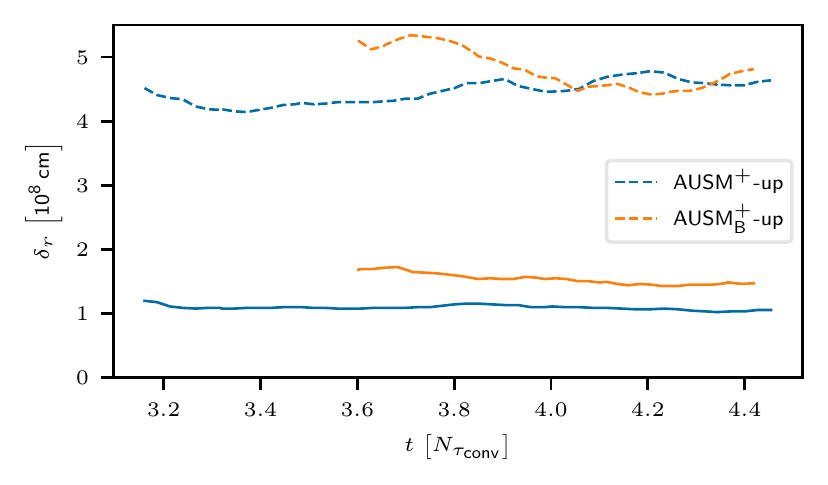}
  \caption{Same as \cref{fig:bwidth_time} but for a resolution of \vhres and
  a fixed energy boosting of \num{3e4}.}
  \label{fig:bwidth_time_hres}
\end{figure}
This offset is much larger than the small fluctuations of the width for the
lower boundary (\cref{fig:bwidth_time_hres}). However, we note that the
boundaries for the highest resolution runs need some time to adapt to the
increased grid resolution and that boundary widths at early times may still
change, as indicated in \cref{fig:bwidth_time}. Therefore, larger time
intervals are needed for a stronger statement on the convergence.

\section{Conclusion}
\label{sec:conclusion}

Our study complements the exploration of convective boundary mixing in stellar
interiors with multidimensional hydrodynamic simulations of convective
helium-shell burning. The initial stratification is based on an 1D \mesa model
of a \SI{25}{\msol} star. \citet{gilkis2016a} use the \mbox{MAESTRO} code to perform
hydrodynamic simulations of convective helium-shell burning in a
\SI{15}{\msol} star. Their study, however, focuses on the angular momentum
distribution within the convection zone and the boundaries to stable layers
above and below the convective shell are not analyzed in detail.

Our 2D and 3D hydrodynamic simulations in spherical-wedge geometry are
performed with the time-implicit, finite-volume Seven-League Hydro (\slh) code.
We calculate the hydrodynamic fluxes with the low-Mach \ausmpup scheme in
combination with \leroux\wbing. Because previous \slh simulations with this
combination had shown that convection is represented properly only for Mach
numbers above \num{e-3}, the energy input had to be boosted to increase the
velocities. We applied boosting factors ranging from \num{3e3} to \num{3e4}. This
results in Mach numbers ranging from $\sim\num{5e-3}$ to $\sim\num{1e-2}$. The
employed grid resolutions range from \lres cells up to \vhres cells.

In order to assess the performance of the \ausmpup solver, we compare different
diagnostic values to a variant of this scheme, denoted as \ausmp, that does not
employ the improved low-Mach capabilities. The flow morphology of fully
developed convection at a resolution of \vhres reveals that \ausmpup reproduces
significantly more small-scale structures than the \ausmp scheme
(\cref{fig:pcolor_flopat}). This is confirmed by the corresponding kinetic
energy spectrum (\cref{fig:spechres}). The spectrum obtained with the \ausmpup
scheme has an inertial range that reaches down to scales a factor of two
smaller than in the spectrum for the \ausmp scheme. The numerical dissipation
as obtained from the \rans kinetic energy equation shows an improved behavior
at the bottom boundary and indicates that the dissipation is converged already
at rather low resolution (\cref{fig:line_rans_b3e4}). For the \ausmp solver,
convergence is found only at the highest resolution of \vhres cells. These
results indicate that a low-Mach method is beneficial already at moderate Mach
numbers. In a future study, simulations of convection with the \ausmpup solver
will be compared in detail to more widely used approaches, as, for example, to
the PPM method.

We analyzed the entrainment rate at the boundaries of the convection zone in
terms of the bulk Richardson number (\cref{eq:rib}). For this, a series of
simulations with varying boosting strength has been carried out on grids with
\lres cells. We found an exponent of $n=0.76$ (\cref{fig:line_bpos_scaling})
which is compatible with \ci and \cii but smaller than results reported for
example by \citet{meakin2007b} or \citet{andrassy2020a} who find $n\approx 1$.
Furthermore, in our simulations a considerable fraction of the measured
entrainment velocity may be attributed to entropy increase in the convection
zone due to the energy release. This is an important aspect if the results of
entrainment studies from hydrodynamic simulations are to be used in 1D
calculations.  Recently, the Bulk-Richardson entrainment scaling was applied to
stellar evolution calculations \citep{staritsin2013a,scott2021a}.
\citet{scott2021a} show that it naturally leads to a mass-dependent efficiency
of \cbm, which is suggested by observations. However, their study indicates
that values of $n<1$ result in a too efficient mixing and that the values for
$A$ that are commonly found in hydrodynamic simulations are too large. Future
simulations, especially at nominal luminosity, may help to identify the origins
of this discrepancy, also regarding the question whether it is applicable only
to a subset of convection zones during stellar evolution as suggested by
\citet{viallet2015a}.

Measuring the widths of the transitions from the convection zone to the
adjacent stable zones showed that the transition is wider for the \lres
resolution compared with simulations on finer grids. This indicates that our
results may not be numerically converged and that our higher-resolution
simulations need to be continued to verify the robustness of our result for the
entrainment rate.  We further assessed the relation between shear strength and
mixing events in our simulations and found that mixing occurs not in the
regions of strongest shear but rather at lower values in the range of measured
shear strengths (\cref{fig:pcolor_shear_mix_low}). This is consistent with the
findings of \citet{woodward2015a}.

Our study has demonstrated that the low-Mach \ausmpup solver is suitable to
address setups that base on realistic stellar models if \wbing is used.
Recently, the Deviation \wbing scheme of \citet{berberich2021a} was added to
the \slh code. In simplified convective test simulations presented by
\citet{edelmann2021a}, the achieved Mach numbers reach $\ma\approx\num{e-4}$.
These velocities are in the regime of convective velocities predicted by \mlt
in early evolutionary phases of stars. The combination of the new Deviation
\wbing method and the \ausmpup solver is a promising approach for future \slh
simulations of stellar convection in the low-Mach regime without the need of
artificially boosted energy generation.

\begin{acknowledgements}
LH, FKR, and RA acknowledge support by the Klaus Tschira Foundation.  The work
of FKR is supported by the German Research Foundation (DFG) through grants
KL~566/22-1 and RO~3676/3-1, respectively. PVFE was supported by the U.S.
Department of Energy through the Los Alamos National Laboratory (LANL). LANL is
operated by Triad National Security, LLC, for the National Nuclear Security
Administration of the U.S. Department of Energy (Contract No.
89233218CNA000001).  The authors gratefully acknowledge the Gauss Centre for
Supercomputing e.V. (www.gauss-centre.eu) for funding this project by providing
computing time through the John von Neumann Institute for Computing (NIC) on
the GCS Supercomputers JUQUEEN \citep{juqueen} and JUWELS \citep{juwels} at
Jülich Supercomputing Centre (JSC). This work also used the DiRAC@Durham
facility managed by the Institute for Computational Cosmology on behalf of the
STFC DiRAC HPC Facility (www.dirac.ac.uk). The equipment was funded by BEIS
capital funding via STFC capital grants ST/P002293/1, ST/R002371/1 and
ST/S002502/1, Durham University and STFC operations grant ST/R000832/1. DiRAC
is part of the National e-Infrastructure.  Furthermore, this article is based
upon work from the “ChETEC” COST Action (CA16117), supported by COST (European
Cooperation in Science and Technology).  RH acknowledges support from the IReNA
AccelNet Network of Networks, supported by the National Science Foundation
under Grant No. OISE-1927130 and from the World Premier International Research
Centre Initiative (WPI Initiative), MEXT, Japan.  This work has been assigned a
document release number LA-UR-21-22119.  The authors thank Cyril Georgy for
useful discussion in the early stages of this project and in particular for
organizing the ISSI Team 367 meeting in 2018.
\end{acknowledgements}

\bibliographystyle{aa}

\begin{appendix}

\section{Reynolds and Favre decomposition}
\label{appendix:reynolds}

The Reynolds decomposition splits a quantity $q(r,\vartheta,\varphi,t)$ in its
mean value $\overline{q}(r)$ averaged over space and time
\begin{align}
  \overline{q}(r) = \frac{1}{\Delta t\Delta\Omega}\int_{\Delta t}\int_{\Delta \Omega} q(r,\vartheta,\varphi,t)\,
                    \dd\Omega\,\dd t,
  \label{eq:rey1}
\end{align}
where $\dd \Omega = \sin\vartheta\,\dd\varphi\,\dd\vartheta$ and the fluctuation
$q'$ is defined as
\begin{align}
  q'(r,\vartheta,\varphi,t) = q(r,\vartheta,\varphi,t) - \overline{q}(r).
  \label{eq:rey2}
\end{align}
The Favre decomposition separates a quantity $q$ into the density-weighted average
\begin{align}
  \tilde{q}(r) = \frac{\overline{\rho q}}{\overline{\rho}}
  \label{eq:fav1}
\end{align}
and the corresponding fluctuation $q''$ defined via
\begin{align}
  q''(r,\vartheta,\varphi,t) = q(r,\vartheta,\varphi,t) - \tilde{q}(r).
  \label{eq:fav2}
\end{align}

\section{Supplementary plots}
\label{sec:appendixplots}

\begin{figure}[h!]
  \centering
  \includegraphics[width=\columnwidth]{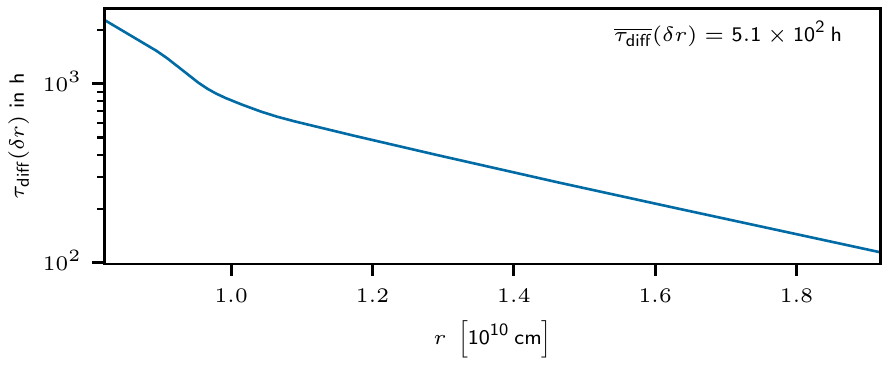}
  \caption{Thermal adjustment timescale $\tdiff$ according to \cref{eq:tdiff}.
  The typical length scale is taken to be the radial grid spacing of the 3D
  simulation run with the highest resolution presented in
  \cref{sec:3Dresults}.}
  \label{fig:line_tdiff}
\end{figure}
\begin{figure*}
  \centering
  \includegraphics[width=\textwidth]{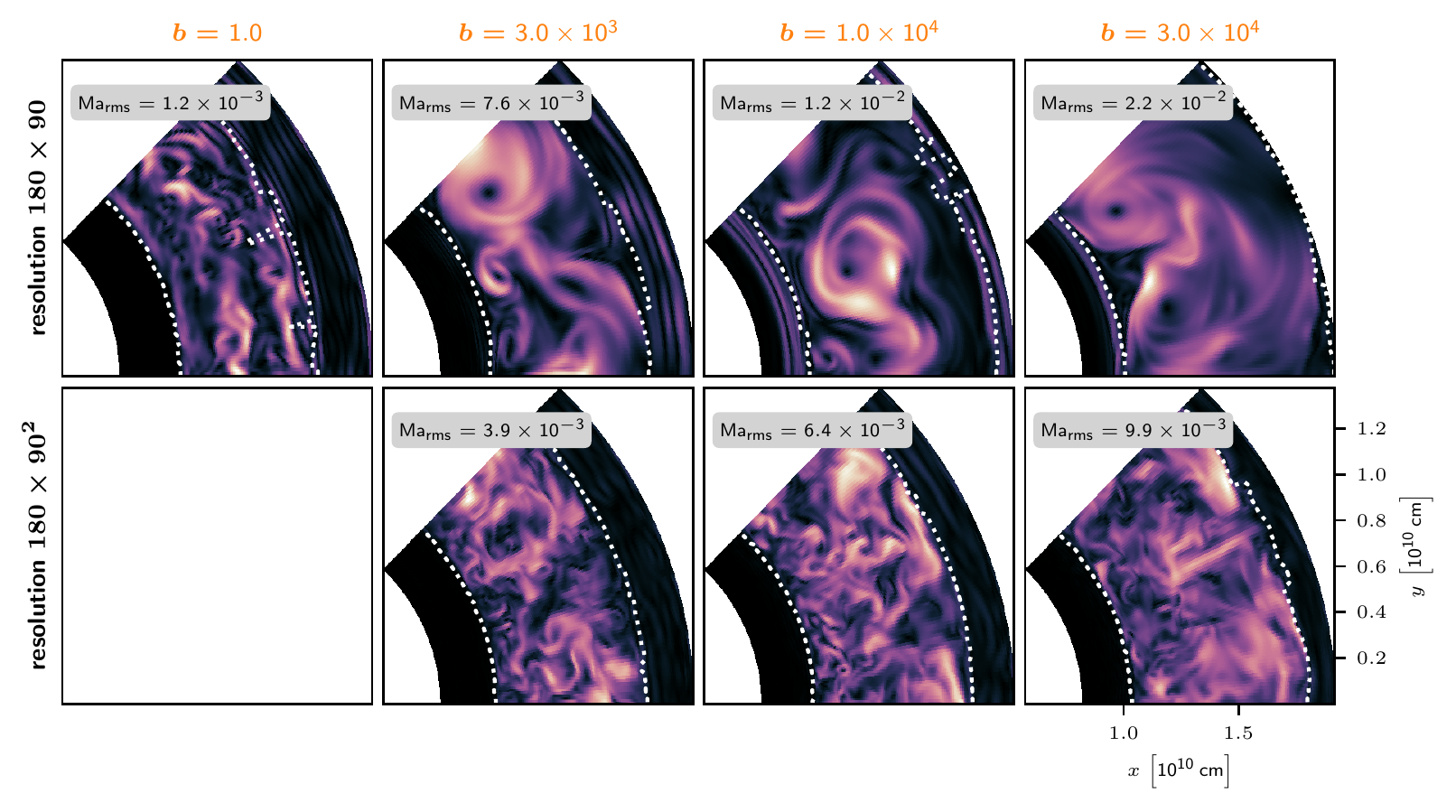}
  \caption{Flow patterns for 2D (upper row) and 3D (lower row) simulations at
  different boosting strengths. The Mach number is color-coded; the color-scale
  is adjusted to every subplot individually. Plots for the 3D simulations show
  the equatorial plane. White dashed lines denote the detected boundaries as
  described in \cref{sec:pstrace}. For all simulations, the \ausmpup solver was
  used. No 3D data is available for $b=1.0$.}
  \label{fig:mach_pcolor_2D_3D}
\end{figure*}
\begin{figure*}
  \centering
  \includegraphics[width=\textwidth]{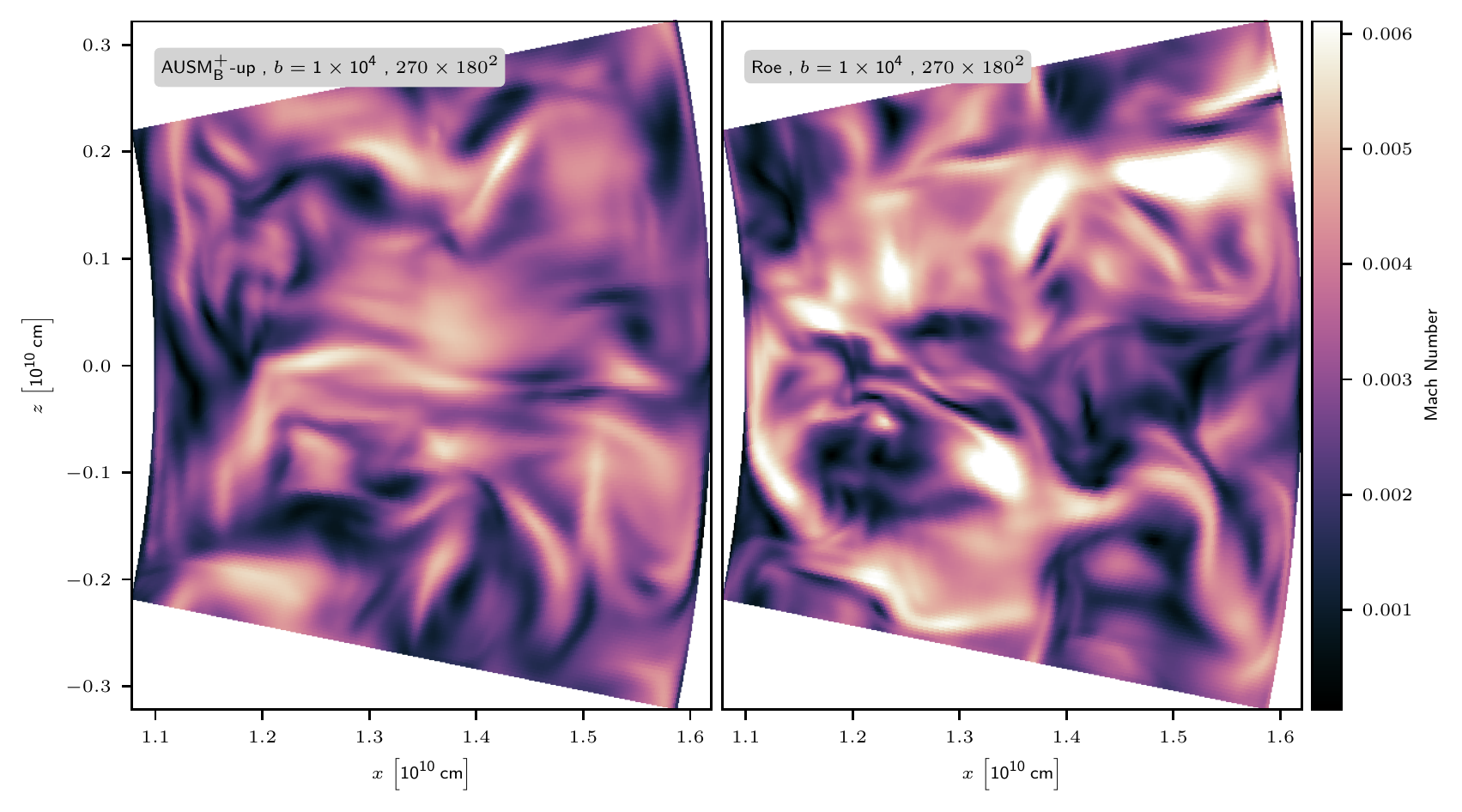}
  \caption{Fluid flow in terms of Mach number for the \ausmp and \roe solver for
  a restricted domain that only contains a fraction of the convection zone.}
  \label{fig:mach_pcolor_roe_ausm}
\end{figure*}
\begin{figure*}
  \centering
  \includegraphics[width=\textwidth]{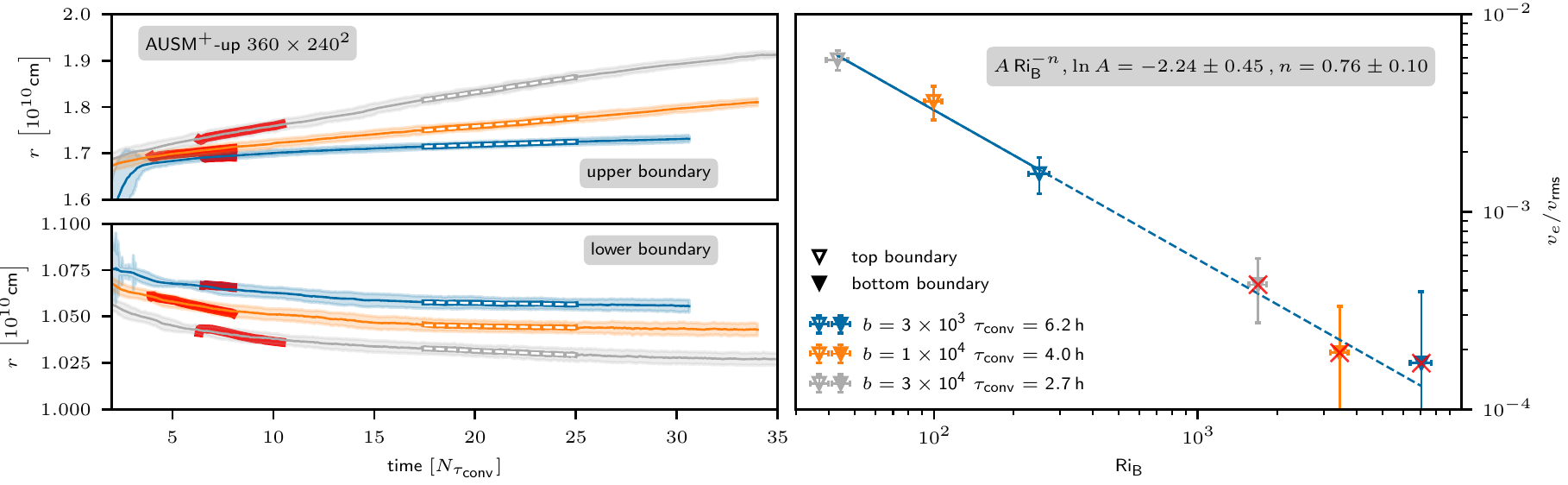}
  \caption{Same as \cref{fig:line_bpos_scaling} but for the time interval $t\in\left[\tofn{17.5},\tofn{25}\right]$.}
  \label{fig:line_bpos_scaling_compare}
\end{figure*}
\clearpage
\end{appendix}
\end{document}